\documentclass[%
 aip,
 php,
 amsmath,amssymb,
 reprint,%
]{revtex4-1}

\draft % marks overfull lines with a black rule on the right

\usepackage{graphicx}
\usepackage{hyperref}
\usepackage{multirow}

\begin{document}

%\preprint{}

\title{Formation of collisionless shocks in magnetized plasma interaction with kinetic-scale obstacles} %Title of paper



\author{F. Cruz}
\email[]{fabio.cruz@tecnico.ulisboa.pt}
%\thanks{}
\affiliation{GoLP/Instituto de Plasmas e Fus\~{a}o Nuclear, Instituto Superior T\'{e}cnico, Universidade de Lisboa, 1049-001 Lisboa, Portugal}

\author{E. P. Alves}
%\email[]{}
%\thanks{}
%\altaffiliation{SLAC National Accelerator Laboratory, Menlo Park, CA 94025, USA}
\affiliation{GoLP/Instituto de Plasmas e Fus\~{a}o Nuclear, Instituto Superior T\'{e}cnico, Universidade de Lisboa, 1049-001 Lisboa, Portugal}
\affiliation{SLAC National Accelerator Laboratory, Menlo Park, CA 94025, USA}

\author{R. A. Bamford}
%\email[]{}
%\thanks{}
\affiliation{RAL Space, STFC, Rutherford Appleton Laboratory, Harwell Oxford, Didcot OX11 0QX, UK}

\author{R. Bingham}
%\email[]{}
%\thanks{}
%\altaffiliation{University of Strathclyde, Glasgow, Scotland, UK}
\affiliation{RAL Space, STFC, Rutherford Appleton Laboratory, Harwell Oxford, Didcot OX11 0QX, UK}
\affiliation{University of Strathclyde, Glasgow, Scotland, UK}

\author{R. A. Fonseca}
%\email[]{}
%\thanks{}
%\altaffiliation{GoLP/Instituto de Plasma e Fus\~{a}o Nuclear, Instituto Superior T\'{e}cnico, Universidade de Lisboa, 1049-001 Lisboa, Portugal}
\affiliation{GoLP/Instituto de Plasmas e Fus\~{a}o Nuclear, Instituto Superior T\'{e}cnico, Universidade de Lisboa, 1049-001 Lisboa, Portugal}
\affiliation{DCTI/ISCTE-Instituto Universit\'{a}rio de Lisboa, 1649-029 Lisboa, Portugal}

\author{L. O. Silva}
\email[]{luis.silva@ist.utl.pt}
%\thanks{}
\affiliation{GoLP/Instituto de Plasmas e Fus\~{a}o Nuclear, Instituto Superior T\'{e}cnico, Universidade de Lisboa, 1049-001 Lisboa, Portugal}


\date{\today}

\begin{abstract}
We investigate the formation of collisionless magnetized shocks triggered by the interaction between magnetized plasma flows and miniature-sized (order of plasma kinetic-scales) magnetic obstacles resorting to massively parallel, full particle-in-cell simulations, including the electron kinetics. The critical obstacle size to generate a compressed plasma region ahead of these objects is determined by independently varying the magnitude of the dipolar magnetic moment and the plasma magnetization. We find that the effective size of the obstacle depends on the relative orientation between the dipolar and plasma internal magnetic fields, and we show that this may be critical to form a shock in small-scale structures. We study the microphysics of the magnetopause in different magnetic field configurations in 2D and compare the results with full 3D simulations. Finally, we evaluate the parameter range where such miniature magnetized shocks can be explored in laboratory experiments.
\end{abstract}

 % PACS numbers


\maketitle %\maketitle must follow title, authors, abstract and \pacs

%\section{}
%\label{}
%\subsection{}
%\subsubsection{}

\section{Introduction}
\label{sec:intro}

The interaction between plasmas and magnetic obstacles is a problem of interest in both space and laboratory plasmas. In general, this interaction is purely three dimensional, highly nonlinear and may happen over a wide range of parameters describing the plasma (\textit{e.g.} magnetization, bulk flow velocity, impact angle) and the obstacle. The complexity of the problem thus limits the development and application of analytical models and requires the use of computer simulations.

In space, the interaction between plasmas and planetary-scale magnetic obstacles leads to the formation of magnetospheres when the magnetic pressure exceeds the plasma pressure, which shield the surface of the planets from energetic particles~\cite{Russel:1991}. From this interaction, a compressed plasma region generally arises as a result of counterstreaming plasmas in the form of a bow shock. For the counterstreaming to occur, it is critical that the plasma is effectively reflected. This may not be the case if the magnetic obstacle is of the order or smaller than the plasma kinetic scales (\textit{i.e.} the ion skin depth and/or gyroradius), even though some particles can be deflected, leading to the formation of a so called mini magnetosphere.

Interest has recently risen in the study of mini magnetospheres, mainly motivated by the observation of crustal magnetic anomalies on the lunar surface~\cite{Colburn:1967, Lin:1998}. The Moon does not possess a global magnetosphere and a bow shock like the Earth~\cite{Colburn:1967}. Interestingly, however, it does have localized regions of magnetic field, whose origin is still not clear~\cite{Dwyer:2011}. The magnitude of the lunar surface magnetic field was mapped by the spacecraft Lunar Prospector, which detected surface fields of the order of $10-100$ nT over regions of $100-1000$ km~\cite{Lin:1998}. The typical ion gyration radius around the solar wind magnetic field is, in this region, of the order of $100-1000$ km, \textit{i.e.} it is comparable to the magnetic object's spatial scale. Unlike large scale magnetic obstacles, miniature magnetospheres are extremely sensitive and vulnerable with respect to variations in solar wind pressure and magnetic field direction. These considerations can be extended to other small planets without a global magnetic field like Mars~\cite{Lillis:2013}, as well as to magnetized asteroids or comets~\cite{Russel:1984} of dimensions on the order of the solar wind ion gyro-radius.

Futuristic applications of mini magnetospheres include the concepts of artificial shielding~\cite{Adams:2005, Parker:2006, Bamford:2014} and propulsion~\cite{Winglee:2000} of spacecrafts. The first concerns about protecting the spacecraft and its crew from hazardous radiation in the interplanetary space using an internal dipolar magnetic field created by superconducting coils. The latter focuses on capturing the momentum of the solar wind via a magnetic sail and thus propel the spacecraft.

The plasma and magnetic field conditions of relevance to space and astrophysical magnetospheric dynamics have recently been achieved in the laboratory, enabling the study of these phenomena in controlled experiments~\cite{Brady:2009, Shaikhislamov:2009, Shaikhislamov:2013, Bamford:2012}. The plasma streams used in laboratory-scaled interactions are most commonly generated by focusing high intensity laser beams on a plastic or metallic target. In this process, the target electrons are heated and expand, creating a collective electric field that drags the ions, hence creating a flow of free charged particles. Using this technique, it is possible to produce plasma flows of densities $n_0 \sim 10^{14} - 10^{15}$ $\text{cm}^{-3}$, bulk velocities of $v_0 \sim 10 - 100$ km/s and intrinsic magnetic fields up to $B_\text{IMF} \sim 10^{-1}$ T. With these experiments, it is possible to mimick the relevant physical processes of space and astrophysical scenarios because they feature identical dimensionless parameters. In the case of experiments with magnetized flows, the plasma parameters are scaled such that they have similar Alfv\'{e}nic Mach number to those that occur in realistic scenarios. For the typical solar wind parameters at 1 AU, the Alfv\'{e}nic Mach number is $M_A \sim 1-10$, where $M_A=v_0/v_A$, with $v_A = B_\text{IMF}/\sqrt{4\pi n_0 m_i}$ ($m_i$ is the mass of the plasma ions). %In these conditions, the plasma electrons are magnetized, whereas the ions are unmagnetized.

Recent experiments of plasma streams colliding with magnetic obstacles have focused on the formation of the density cavity. Brady \textit{et al.}~\cite{Brady:2009} studied the macroscopic features of the cavity formation process and observed that the magnetic field pressure that balances the plasma ram pressure could be accurately estimated from the magnetohydrodynamics (MHD) formalism. In other works~\cite{Shaikhislamov:2009, Shaikhislamov:2013}, the role of smaller scale physics at the boundary between the plasma and the density cavity was addressed, and including the Hall current in the MHD framework was found to be consistent with an asymmetry on the overall shape of that boundary observed experimentally. More recently, Bamford \textit{et al.}~\cite{Bamford:2012} also studied miniature systems experimentally, by using a solar wind tunnel to generate a collisionless, supersonic flow that collided against the dipolar magnetic field of a magnet. In this work, the conditions for the formation of mini magnetospheres in laboratory scenarios were investigated, confirming a previous numerical study by Gargat\'{e} \textit{et al.}~\cite{Gargate:2008}. In all these experimental works, the plasma streams presented a non-negligible degree of collisionality. However, recent progress in achieving collisionless conditions in the laboratory~\cite{Kugland:2012, Fox:2013, Niemann:2014, Huntington:2015, Schaeffer:2016} opened the possibility to perform experimental studies of kinetic-scale collisionless physics relevant in astrophysical scenarios.

Previous numerical approaches to the problem, particularly focused on the interaction between the solar wind and lunar magnetic anomalies, used mostly magnetohydrodynamic (MHD) and hybrid simulations. Using MHD simulations, Harnett and Winglee~\cite{HarnettWinglee:2000, HarnettWinglee:2002, HarnettWinglee:2003} found that mini magnetospheres show strong variations in size and shape depending on the interplanetary magnetic field orientation. Nevertheless, they identified regions where ion and electron particle dynamics (not resolved in the MHD approach) might be important, namely regions close to where the reflection of the solar wind occurs, usually called magnetopause. More recent hybrid simulations confirmed the importance of kinetic effects in these systems: Gargat\'{e} \textit{et al.}~\cite{Gargate:2008} modelled the collision of a plasma flow with a magnetic dipole using hybrid simulations with realistic parameters. Besides showing good qualitative agreement with experimental results, their work included a simple model to evaluate the pressure balance at the magnetopause which was verified for different plasma conditions. Gargat\'{e} \textit{et al.}~\cite{Gargate:2014} have also used hybrid simulations to study the formation of a shock driven by a coronal mass ejection, following the shock evolution on the ion time scale and identifying purely kinetic effects such as ion acceleration. Finally, Blanco-Cano \textit{et al.}~\cite{Blanco-Cano:2004} used hybrid simulations of plasma flows interacting with dipolar obstacles to show that a magnetosphere is only formed if the obstacle size is much larger than the ion inertial length.

Correctly modelling these systems implies understanding the kinetic-scale phenomena of the plasma. Particle-in-cell (PIC) simulations play a critical role in this effort since they can capture the important microphysical processes underlying the formation of small-scale magnetospheres. Only recently full particle simulations were used to model directly a lunar magnetic anomaly~\cite{Deca:2014, Deca:2015}. In these works, Deca \textit{et al.} showed that electron dynamics dominates the near-surface plasma environment. In particular, this work showed not only that non-Maxwellian particle distributions are generated from the interaction with the mini magnetosphere, but also that the plasma deflection occurs due to microscopic collective electric fields associated with charge separation between electrons and ions, which can only be appropriately captured using PIC simulations. Ashida \textit{et al.}~\cite{Ashida:2014} studied the interaction between an unmagnetized plasma flow and magnetic obstacles with sub-Larmor radius magnetic obstacles, showing that mini magnetospheres can be formed even for obstacles sizes smaller than the ion gyroradius. Other recent works~\cite{Bamford:2012, Bamford:2016} with full PIC simulations show that enhanced proton flux around lunar magnetic anomalies can be responsible for the appearance of dark lanes on lunar swirls, regions on the lunar surface commonly found around mini magnetospheres that receive enhanced ageing from direct interaction with the solar wind.

None of the previous studies have identified the formation of collisionless shocks in these plasma interactions with miniature obstacles. It is expected that there should be a critical obstacle size above which the formation of a collisionless shock should occur, similarly to what occurs in planetary scales. Therefore the conditions for the formation of collisionless shocks and the transition between shock-forming and non-shock-forming obstacles remains to be addressed via first principle simulations. In this work, we model the interaction between a magnetized plasma colliding with a dipolar magnetic field using multidimensional PIC simulations, focusing on obstacles with sizes comparable to the plasma kinetic scales.

This paper is organized as follows. In Sections~\ref{sec:constantrhoi_varyingL} and \ref{sec:constantL_varyingrhoi}, we show that the formation of shocks in mini magnetospheres is critically determined by the ratio between the obstacle size and the ion Larmor radius (determined by the plasma magnetization) by independently controlling the magnitude of the dipole moment and the plasma magnetization. In Section~\ref{sec:IMForientation}, we show that the effective obstacle size is, in the case of small-scale obstacles, dependent on the relative orientation between the dipolar and the plasma internal magnetic fields. We develop an analytical model for the effective obstacle size and show that the inflation/deflation of the cavity may, in some cases, be critical to observe shock formation. In Section~\ref{sec:inplane}, we qualitatively discuss the effects of field-aligned dynamics in the magnetopause structure. The importance of 3D interplay effects in cavity and shock properties is assessed in Section~\ref{sec:3Dinterplay}. In Section~\ref{sec:labparams}, we use the results presented in Sections~\ref{sec:constantrhoi_varyingL}-\ref{sec:3Dinterplay} to evaluate the possibility of generating collisionless shocks in mini magnetospheres in laboratory and space scenarios. Recent experimental results are interpreted and parameters for future experiments are discussed. Finally, we state the conclusions in Section~\ref{sec:conclusions}.

\section{Shock formation in mini magnetospheres}
\label{sec:shockform}

In order to accurately model the interaction between a plasma flow and a small-scale obstacle, full PIC simulations are critical due to the intrinsically kinetic character of the system. In this work, we present simulations performed using OSIRIS~\cite{Fonseca:2012, Fonseca:2013}, a massively-parallel and fully relativistic PIC code. Using OSIRIS, we are able to capture high frequency phenomena, as well as kinetic-scale physics such as finite Larmor radius effects and non-Maxwellian particle distributions. 
 
OSIRIS operates in normalised plasma units, the independent variable being the plasma density $n_0$. Distances are normalised to the electron skin depth $d_e \equiv c/\omega_{pe}$ (where $c$ is the speed of light and $\omega_{pe} = \sqrt{4\pi n_0 e^2/m_e}$ is the plasma frequency, with $e$ and $m_e$ representing the electron charge and mass, respectively) and times are normalised to the inverse of the plasma frequency $1/\omega_{pe}$.

In all the simulations, we use a cold plasma stream with fluid velocity $v_0 = 100 v_{the}$, where $v_{the}$ is the electron thermal velocity. Although finite plasma temperatures will certainly play a role in the structure of the generated shocks, we neglect thermal effects in this first approach in order to simplify the analysis. For computational purposes, we use a reduced ion-to-electron mass ratio $m_i/m_e = 100$. This parameter controls the separation between ion and electron temporal and spatial scales, and was chosen such that no significant changes are observed in the simulation results when compared to test simulations with approximately half the realistic ratio ($m_i/m_e = 900$). By using these parameters, we can significantly reduce the computational effort to perform the numerical experiments and yet are still able to gain important physical insight into the dynamics of these complex systems. The simulation domain is filled with the plasma internal magnetic and electric fields $\mathbf{B_\text{IMF}}$ and $\mathbf{E_\text{IMF}}$ such that $\mathbf{E_\text{IMF}} + \mathbf{v_0} \times \mathbf{B_\text{IMF}} = 0$. The magnitude of $\mathbf{B_\text{IMF}}$ is chosen such that the flow has a given $M_A = v_0 \sqrt{4\pi n_0 m_p} /B_\text{IMF}$. A dipolar magnetic field is externally imposed. The dipole moment $m$ is chosen such that the plasma ram pressure equals the magnetic pressure associated with a magnetic field $B_\text{RMP}$, measured at a distance $L_0$ from the dipole which is comparable to the plasma kinetic scales. The MHD pressure balance reads, at this point,
\begin{equation}
n_0 m_i v_0^2 = \frac{B_\text{RMP}^2}{8\pi} \text{ ,} \ \ \ \ B_\text{RMP} = \frac{m}{L_0^3} \text{ .}
\label{eq:MHDpbalance}
\end{equation}
This macroscopic picture has an underlying, well understood microscopic equivalent~\cite{Deca:2014, Bamford:2016}. As the plasma approaches the steep gradient of the magnetic field at the magnetopause, it is slowed down due to a ponderomotive-like force ($\nabla B^2$). Due to their different inertia, the penetration depth of ions and electrons in this region is slightly different. Thus, a collective, charge separation electric field is set up at the magnetopause, which is responsible for deflecting/reflecting the incoming plasma particles. In the case of finite magnetic obstacles, we shall observe both dynamics for plasma ions: specular reflection is expected if the ions collide with the central region of the obstacle, and mere deflection if they collide with the flanks of these obstacles.
% add paragraph on the microphysics of the deflection/standoff. Talk about the layer of electric field and the “specular reflection” from it.

In the sections below, we show simulation results for different plasma and dipole conditions. In particular, we study the formation of magnetized shocks in mini magnetospheres by varying the plasma parameters and the dipolar moment independently (Sections~\ref{sec:constantrhoi_varyingL} and \ref{sec:constantL_varyingrhoi}, respectively).

\subsection{Critical obstacle size for fixed ion gyroradius}
\label{sec:constantrhoi_varyingL}

\begin{figure}
\includegraphics[height=6.83cm]{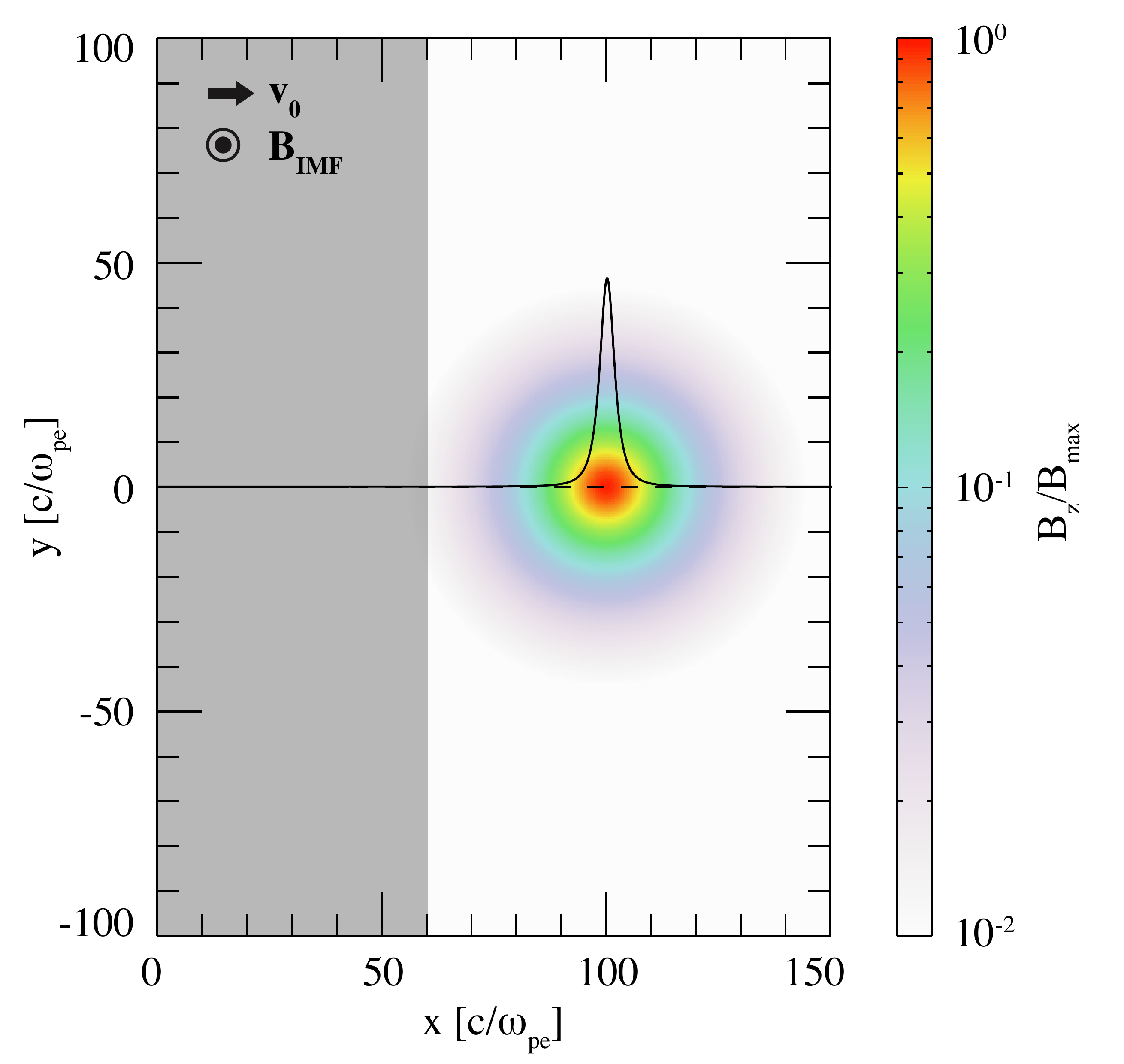}
\caption{\label{fig:setup_Bz} Schematic illustration of the initial setup of the 2D simulations. The plasma is continuously injected from the left boundary and flows along the $x$ direction. Part of the plasma is initialized inside the box (indicated in grey), to reduce the computation effort. The colour map indicates the dipole field strength normalized to its maximum value, determined by a cutoff introduced in the dipolar field. A lineout (at $y=0$) of this field is also indicated.}
\end{figure}

We first start by addressing the critical obstacle size that allows shock formation in miniature obstacles. We consider the interaction between a plasma of $v_0 = 0.2c$ and $M_A = 2$ and a dipolar magnetic field of increasing magnitude.  Although the plasma flow chosen here is much faster than typical space plasma flows, the physics that we are focused on depends on $M_A$ (as shown below), and therefore we are simply scaling up the system for computational purposes. The same results have been verified for the same $M_A$ with lower fluid velocities, up to $v_0=0.02c$. A schematic illustration of the initial setup of the 2D simulations presented below is shown in Fig.~\ref{fig:setup_Bz}. The plasma is continuously injected from the left boundary of the simulation box. Periodic/open boundary conditions are used in the direction perpendicular/parallel to the plasma flow. The simulations are stopped when recirculation occurs in the periodic direction. The grid resolution is 10 cells/$d_e$, with 25 simulation particles per cell per species (electrons and ions). The dipolar and the internal plasma magnetic fields are parallel to each other and point out of the simulation plane (positive $z$ direction).

\begin{figure*}
\includegraphics[height=11.5cm]{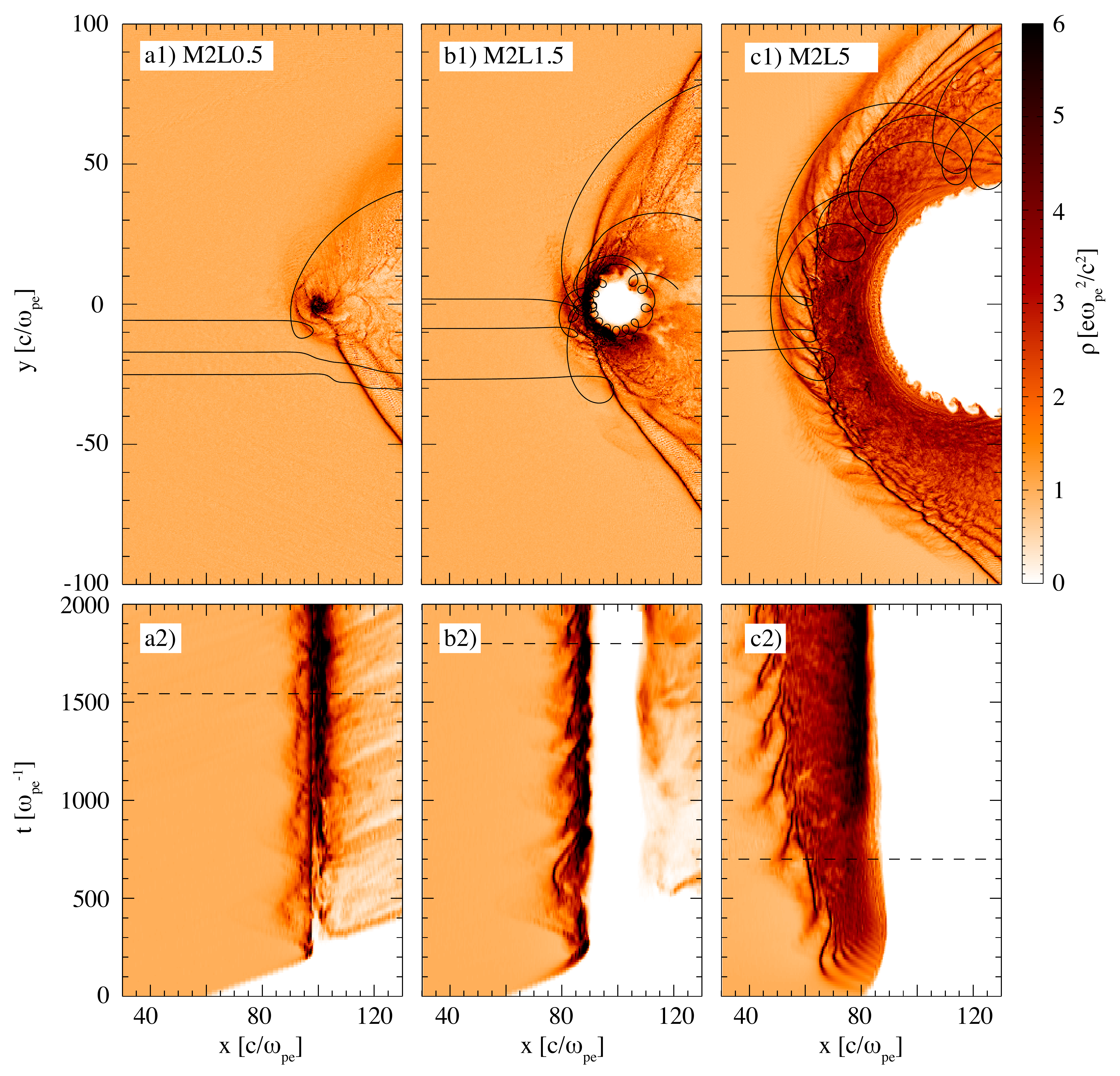}
\caption{\label{fig:bdipolestrength} Critical cavity size for formation of miniature magnetized shocks. A plasma flow with $v_0=0.2c$ and $M_A = 2$ interacts with magnetic dipoles that standoff the plasma at a distance (a) $L_0 = 0.25 \rho_i$, (b) $0.75 \rho_i$ and (c) $2.5 \rho_i$. Along with the ions density, black lines are representing ion trajectories are shown in panels a-c1, illustrating the typical Larmor radius scale after the particles are reflected. Panels a-c2 show the time evolution of the same density along the $y=0$ lineout. The dashed lines in these panels indicate the time at which the frames shown above were taken.}
\end{figure*}

In the simulations presented in this section, the dipole magnetic moment was chosen such that the pressure equilibrium in Eq.~(\ref{eq:MHDpbalance}) is satisfied at a distance (a) $L_0 = 0.5 \ d_i$, (b) $1.5 \ d_i$ and (c) $5 \ d_i$ from the dipole, where $d_i = d_e \sqrt{m_i/m_e} $ is the ion skin depth, related with the ion gyroradius $\rho_i$ via the flow Mach number by
\begin{equation}
\rho_i = \frac{m_i c v_0}{e B_\text{IMF}} = M_A d_i \text{ .}
\label{eq:rhoi_MA}
\end{equation}
The parameters used in these simulations are summarised in Table~\ref{tab:simparams}, where a label for each numerical experiment is also given to allow the clear identification of their results in Fig.~\ref{fig:bdipolestrength}.

The plasma flows against and around the dipolar structure and is eventually reflected/deflected when the magnetic pressure equals the plasma ram pressure. In this process, an approximately circular density cavity is created with size (a) $L_0 < d_i, \rho_i$, (b) $d_i < L_0 < \rho_i$ and (c) $d_i, \rho_i < L_0$ (see Fig.~\ref{fig:bdipolestrength}). For $L_0 < d_i, \rho_i$ (Fig.~\ref{fig:bdipolestrength} a1), the dipolar structure can only perturb the plasma creating a wake behind it. After the particles are deflected, they are rotated by the internal plasma magnetic field with a gyroradius $\rho_i > L_0$, \textit{i.e.} the plasma is not able to pile up in front of the cavity size and create a compressed (shocked) region of magnetic field. A similar result is observed for $d_i < L_0 < \rho_i$ (Fig.~\ref{fig:bdipolestrength} b1), even though compressed plasma regions show an oscillatory dynamics ahead of the magnetic obstacle. The plasma ions that are specularly reflected on the magnetic obstacle form a structure similar to a shock foot (as seen in Fig.~\ref{fig:bdipolestrength} c1), but they flow around the cavity as their gyroradius is also larger than the obstacle size. In the case where $L_0 > d_i, \rho_i$ (Fig.~\ref{fig:bdipolestrength} c1), the plasma ions can be reflected in front of the magnetic obstacle and thus counterstream with the unperturbed flow, leading to the generation of turbulence via the modified two-stream instability~\cite{McBride:1972}. A curved shock front, clearly identified by the sharp transition between the unperturbed and compressed  plasma regions, is formed ahead of the density cavity. The latter region is typically called the magnetosheath and is characterised by its turbulent structures. The results presented here suggest that the critical kinetic-scale requirement that determines the shock formation in mini magnetospheres is $L_0/\rho_i > 1$. The difference between the oscillatory dynamics discussed above and the formation of a shock can be observed from Figs.~\ref{fig:bdipolestrength} a-c2, where the time evolution of the ion density along the $y=0$ lineout is shown. Whilst we observe the jump in density oscillating back and forth in time for $L_0 < \rho_i$ (Figs.~\ref{fig:bdipolestrength} a2 and b2), it is clear that the compressed plasma region increases in time for $L_0 > \rho_i$ (Fig.~\ref{fig:bdipolestrength} c2). Throughout this work, we consider a shock is formed when the plasma is continuously compressed ahead of the obstacle, as illustrated in Figs.~\ref{fig:bdipolestrength} c1-2.

\begin{table}
\caption{\label{tab:simparams} Parameters of the simulations presented in Sections~\ref{sec:constantrhoi_varyingL} and \ref{sec:constantL_varyingrhoi}. The labels shown here can be used to identify each simulation in Figs.~\ref{fig:bdipolestrength} and \ref{fig:flowMA_zdipole}.}
\begin{ruledtabular}
\begin{tabular}{llllll}                          
Sim. group & Label & $M_A$ & $L_0/d_i$ & $L_0/\rho_i$ & Relevant ordering \\ \hline
\multirow{3}{*}{Section~\ref{sec:constantrhoi_varyingL}} & M2L0.5 & 2 & 0.5 & 0.25 & $L_0 < d_i, \rho_i$ \\
 & M2L1.5 & 2 & 1.5 & 0.75 & $d_i < L_0 < \rho_i$ \\
 & M2L5 & 2 & 5 & 2.5 & $d_i, \rho_i < L_0$ \\ \hline
\multirow{3}{*}{Section~\ref{sec:constantL_varyingrhoi}} & M1.5L2 & 1.5 & 2 & 1.33 & $d_i,\rho_i < L_0$ \\
 & M3L2 & 3 & 2 & 0.67 & $d_i < L_0 < \rho_i$ \\
 & M10L2 & 10 & 2 & 0.2 & $d_i < L_0 < \rho_i$ \\
\end{tabular}
\end{ruledtabular}
\end{table}

We also note that, in all the cases, microscopic instabilities are developed at the magnetopause, due to a relative ion-electron drift. The origin of such drift will be discussed in Section~\ref{sec:IMForientation}.

\subsection{Critical ion gyroradius for fixed obstacle size}
\label{sec:constantL_varyingrhoi}

\begin{figure*}
\includegraphics[height=11.6cm]{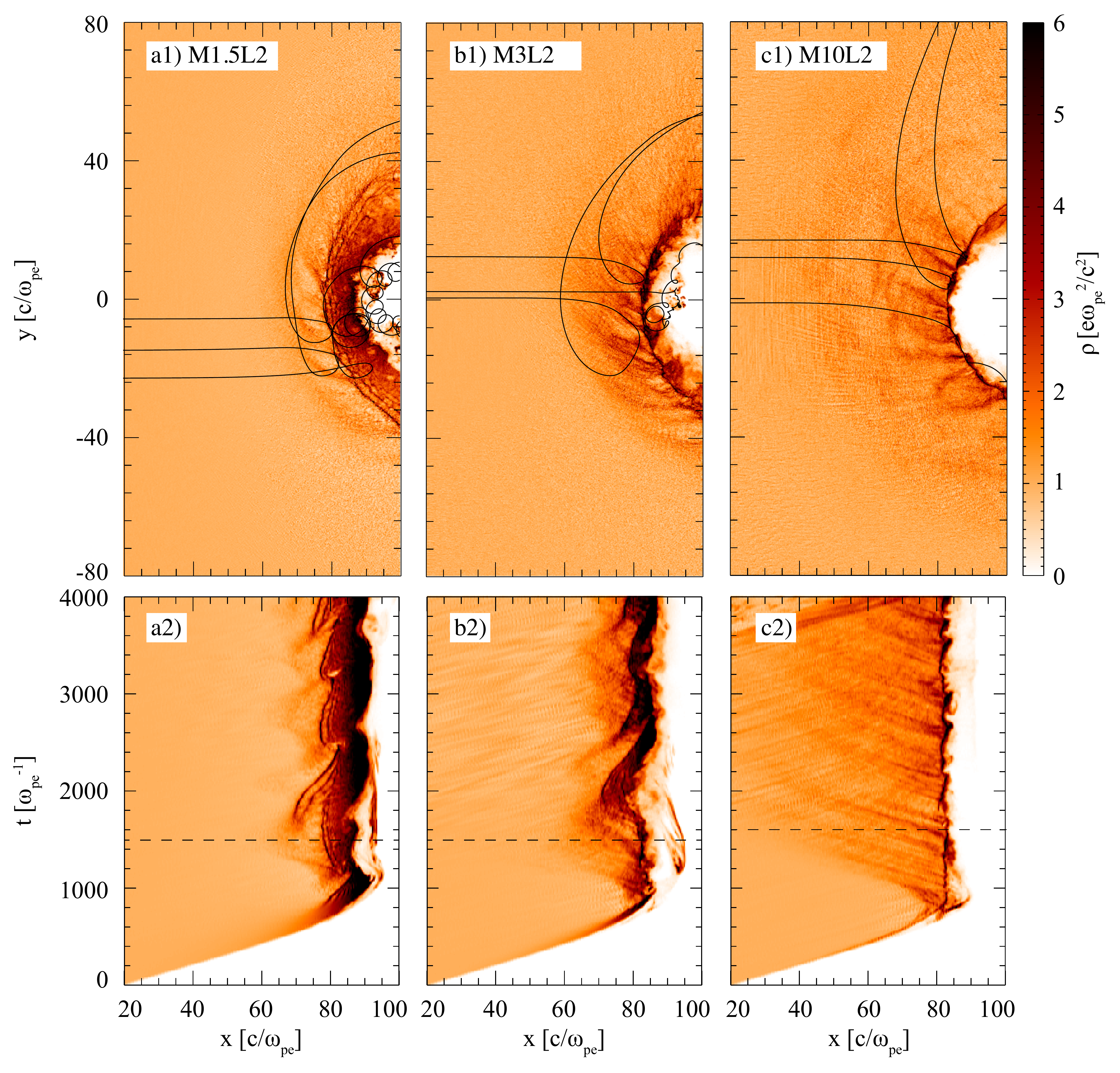}
\caption{\label{fig:flowMA_zdipole} Critical ion gyroradius for fixes obstacle size. Three plasma flows with (a) $M_A = 1.5$, (b) $M_A = 3$ and (c) $M_A = 10$ interact with a magnetic dipole that standoff the plasma at a distance $L_0 = 2 d_i$. Panels a-c2 show the time evolution of the ion density along the $y=0$ lineout. The dashed lines in these panels indicate the time at which the frames shown above were taken.} 
\end{figure*}

Let us now consider the interaction between a constant dipolar field and plasma flows with different Mach numbers. Since the ion gyroradius scales with $M_A$ according to Eq.~(\ref{eq:rhoi_MA}), it is also possible to control the ratio $L_0/\rho_i$ by changing the flow $M_A$. We consider a plasma flow with $v_0 = 0.1 c$ and a dipolar field that holds the plasma ram pressure at $L_0 = 2 d_i$. We consider three flows with (a) $M_A = 1.5$, (b) $M_A = 3$ and (c) $M_A = 10$ (see Fig.~\ref{fig:flowMA_zdipole}). These parameters correspond to (a) $L_0/\rho_i \simeq 1.3$, (b) $L_0/\rho_i \simeq 0.7$ and (c) $L_0/\rho_i \simeq 0.2$. According to the discussion presented in Section~\ref{sec:constantrhoi_varyingL}, only the flow with (a) $M_A = 1.5$ should produce a plasma compressed region, whereas the sub-Larmor-radii obstacles in cases (b) and (c) should not be able to form a shock. The parameters used in these simulations are also summarised in Table~\ref{tab:simparams}.

We observe the formation of a shock for $M_A = 1.5$ (Fig.~\ref{fig:flowMA_zdipole} a) and the same oscillatory dynamics as in Fig.~\ref{fig:bdipolestrength} b) for $M_A=3$ (Fig.~\ref{fig:flowMA_zdipole} b). For $M_A = 10$, the ion gyroradius is much larger than the obstacle size, as the black lines representing ion trajectories in Fig.~\ref{fig:flowMA_zdipole} c) indicate. In this case, the plasma does not develop a shocked region. Similarly to the results of Section~\ref{sec:constantrhoi_varyingL}, these results strongly suggest that a shock can be formed for $L_0/\rho_i > 1$. This condition means that there is a maximum flow $M_A$ for an obstacle with a given size to be able to form a shock. Consequently, the same condition also limits the maximum Alfv\`{e}nic Mach number of the collisionless shocks formed in these interactions.

\subsection{Dependance of the effective obstacle size on IMF orientation}
\label{sec:IMForientation}

\begin{figure}
\includegraphics[height=7.1cm]{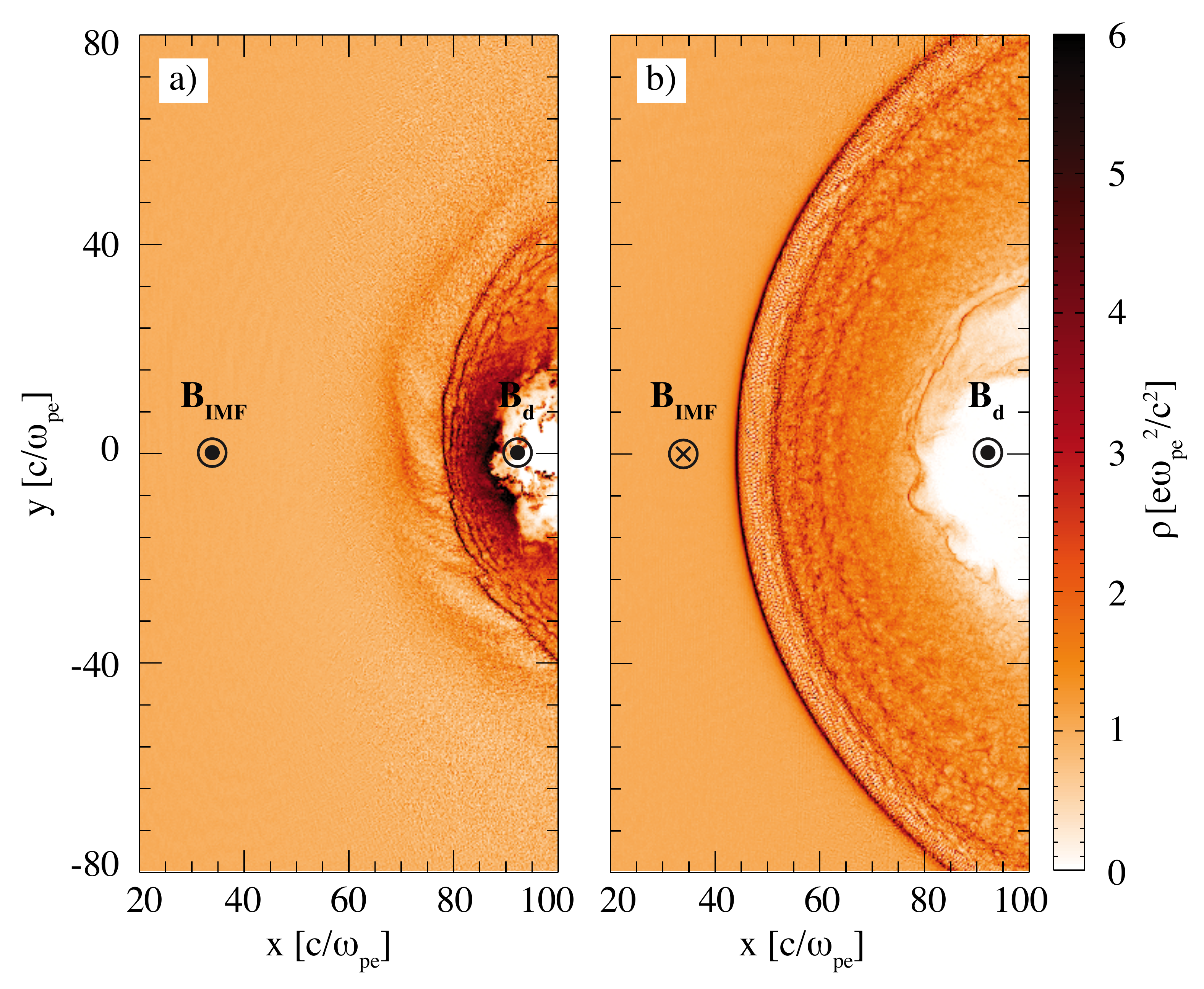}
\caption{\label{fig:bimforientation} Effective cavity size is sensitive to $\mathbf{B_\text{IMF}}$ orientation. A plasma flow with $v_0=0.1c$ and $M_A = 1.5$ is collided with a dipolar magnetic parallel $\mathbf{B_\text{d}}$ (a)/anti parallel (b) to $\mathbf{B_\text{IMF}}$. The plasma is stopped at a distance $L_0 = 2 d_i$ according to the macroscopic pressure balance.}
\end{figure}

On the interaction between the plasma and the magnetized obstacle, the magnetopause position is controlled by the pressure balance. In this subsection, we show that opposite orientations of $\mathbf{B_\text{IMF}}$ can change the total magnetic pressure profile close to the magnetopause and thus inflate or deflate the density cavity. Although these changes may not be relevant in large-scale (\textit{e.g.} planetary) systems, we find that for mini magnetospheres such inflation/deflation can be on the order of 100\% the cavity size for low $M_A$ flows and critically determine the formation of collisionless shocks (see Fig.~\ref{fig:bimforientation}).

\begin{figure}
\includegraphics[height=7.9cm]{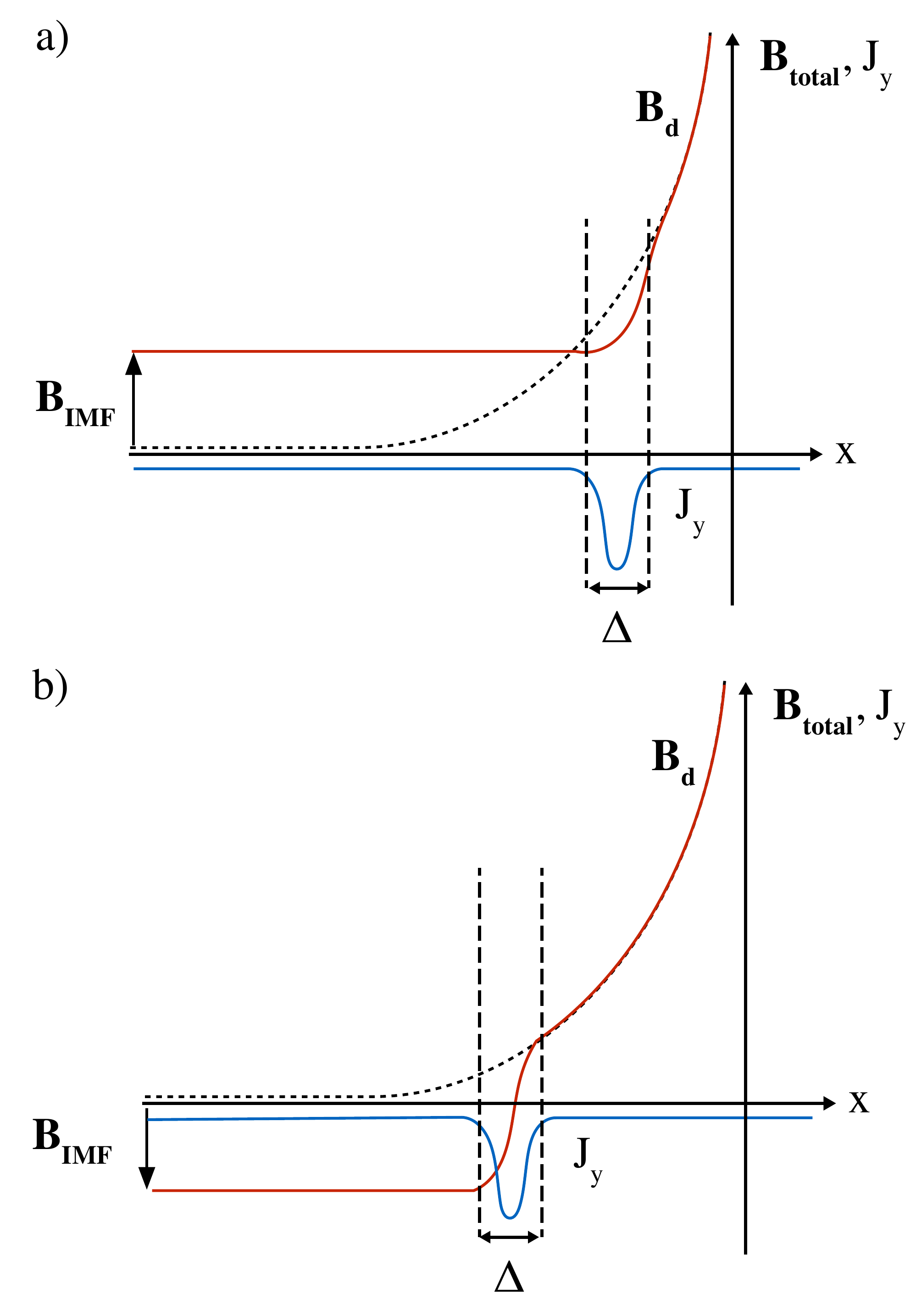}
\caption{\label{fig:physpic} Physical picture of the two-fluid model used to describe the cavity size inflation/deflation in mini magnetospheres depending on the relative orientation between $\mathbf{B_\text{d}}$ and $\mathbf{B_\text{IMF}}$.}
\end{figure}

In general, we find that the magnetic pressure gradient required to stop the plasma flow occurs farther from the dipole for anti parallel $\mathbf{B_\text{d}}$ and $\mathbf{B_\text{IMF}}$ when compared to the opposite relative orientation. This inflation in cavity size can be explained from a simple 1D model based on the physical picture presented in Fig.~\ref{fig:physpic}.

We assume that the magnetopause is stationary (\textit{i.e.} all time derivatives vanish) and only variations aligned with the flow (\textit{i.e.} in the $x$ direction) are allowed. Both $\mathbf{B_\text{d}}$ and $\mathbf{B_\text{IMF}}$ are in the $z$ direction. Since the plasma is highly conductive, surface currents shield the plasma from the magnetic field outside such that its internal magnetic field remains constant. Thus, a gradient in the total magnetic field occurs at the magnetopause, supported by a current $J_y$ that can be estimated using the $y$ component of Amp\`{e}re's law
\begin{equation}
\frac{4\pi}{c} J_y = - \frac{\partial B_z}{\partial x} \text{ .}
\end{equation}
Assuming quasineutrality, the current is given by $J_y = e n_0 (v_{iy} - v_{ey}) \simeq - e n_0 v_{ey}$, where $v_{sy}$ is the velocity along $y$ of species $s$ (with $s=e,i$ corresponds to electrons and ions, respectively). The approximation used above ($v_{iy} \ll v_{ey}$) can be justified by considering the $y$ component of the momentum equation describing the electron and ion fluids, 
\begin{equation}
\label{eq:momentumyy}
v_{sx} \frac{\partial}{\partial x} v_{sy} = \frac{q_s}{m_s}\left( E_y - \frac{v_{sy}}{c} B_z \right) \text{ .}
\end{equation}
Combining both species' equations and assuming $v_{ex} \simeq v_{ix}$ (once again due to quasineutrality), we can write
\begin{equation}
v_{iy} = - \frac{m_e}{m_i} v_{ey} \text{ ,}
\end{equation}
\textit{i.e.} the ion velocity along $y$ is much smaller than the electron velocity in the same direction. Hence, the current $J_y$ is driven mainly by an electron drift along $y$. This drift can then be estimated as
\begin{equation}
\label{eq:Bjump}
v_{ey} = - \frac{J_y}{n_0} = \frac{c}{4\pi n_0} \frac{\partial B_z}{\partial x} \simeq \frac{c}{4\pi n_0} \frac{B_\text{d}-B_\text{IMF}}{\Delta} \text{ ,}
\end{equation}
where $\Delta$ is the typical width of the magnetic field jump, which is on the order of $d_e$, as confirmed by the simulation results shown in Fig.~\ref{fig:bimforientation}.
Considering now that the electrons are in equilibrium along the $x$ direction, we can write the electric field along $x$ as
\begin{equation}
\label{eq:EvB}
E_x = - v_{ey}B_z \simeq - v_{ey} B_d \text{ .}
\end{equation}
Combining Eqs.~(\ref{eq:Bjump}) and (\ref{eq:EvB}), we can write the electrostatic potential associated with the electric field $E_x$ as
\begin{equation}
\label{eq:potential}
\phi = - \int E_x \ \mathrm{d}x \simeq - E_x \Delta \simeq \frac{B_\text{d}}{4\pi n_0} (B_\text{d} - B_\text{IMF}) \text{ .}
\end{equation}
This shows that, considering two flows with the same velocity and magnetization and opposite orientations of $B_\text{IMF}$, the electrostatic potential necessary to reflect the plasma occurs for lower $B_\text{d}$ if $B_\text{IMF}$ is negative (\textit{i.e.} if $\mathbf{B_\text{d}}$ and $\mathbf{B_\text{IMF}}$ are anti parallel), as depicted in Fig.~\ref{fig:physpic}. This results in a larger cavity size, as illustrated in the simulations presented in Fig.~\ref{fig:bimforientation}. The plasma reflection occurs when the potential energy density associated with $\phi$ equals the energy density of the incoming flow, \textit{i.e.} when
\begin{equation}
\label{eq:Ebalance}
n_0 \phi = n_0 m_i v_0^2 + \frac{B_\text{IMF}^2}{8\pi} \text{ .}
\end{equation} 

\begin{figure}
\includegraphics[height=5.6cm]{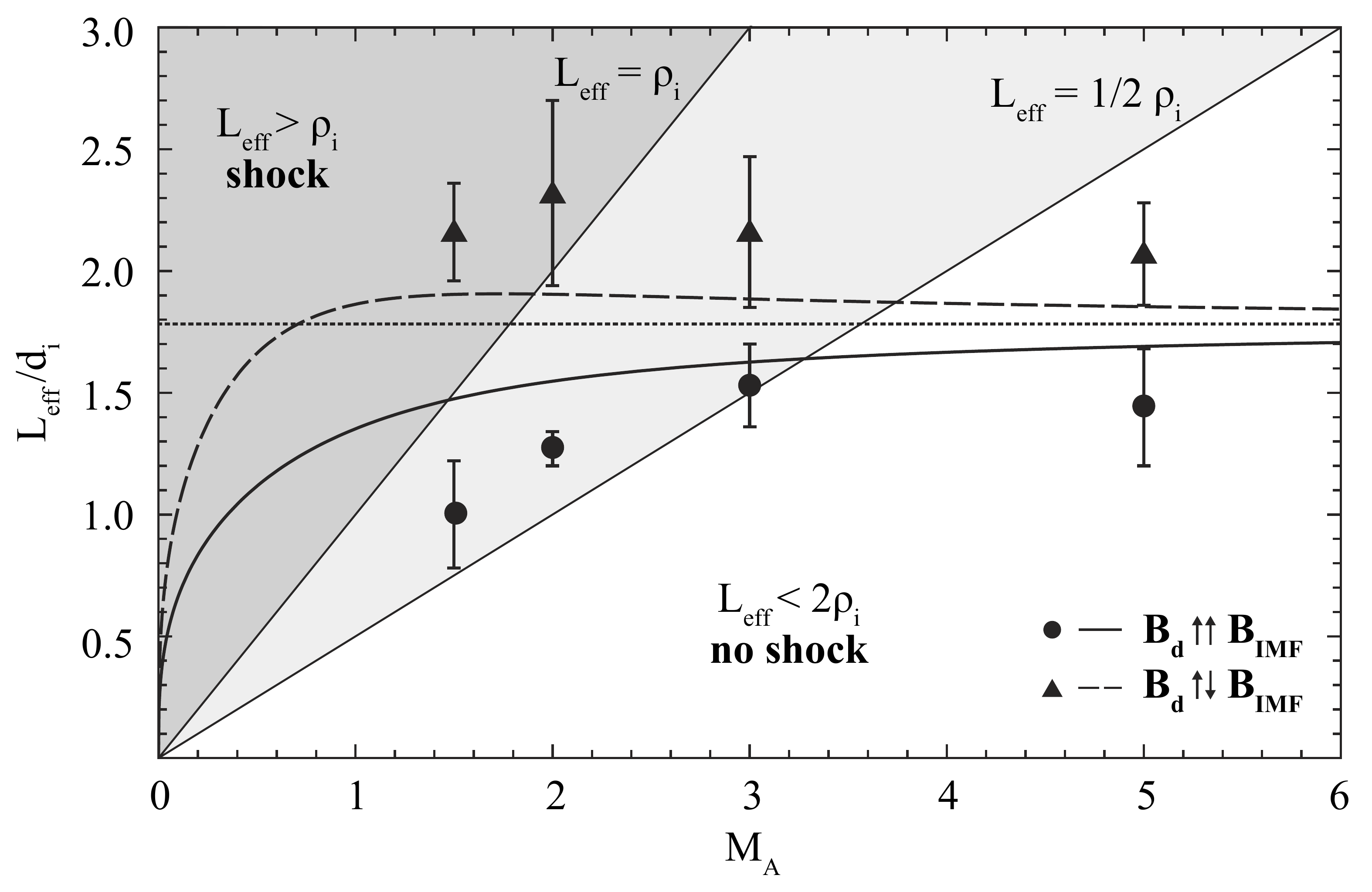}
\caption{\label{fig:2Dcavsize} Effective cavity size as a function of $M_A$ and $\mathbf{B_\text{IMF}}$ orientation. A plasma flow with $v_0=0.1c$ is collided with a dipolar magnetic field that holds the plasma ram pressure at a distance $L_0 = 2 d_i$ according to the macroscopic pressure balance. The estimates for parallel and anti parallel orientations between $\mathbf{B_\text{d}}$ and $\mathbf{B_\text{IMF}}$ are represented in long-dashed and solid lines.}
\end{figure}

In Fig.~\ref{fig:2Dcavsize}, we compare the effective size of a magnetic obstacle with $L_0 = 2 d_i$ estimated with this analytical model with simulation results of a cold plasma flow with constant velocity $v_0 = 0.1 c$ and varying $M_A$ (\textit{i.e.} varying $B_\text{IMF}$). The effective size of the magnetic obstacle is calculated from the magnetic field profile $B_d = m/L_\text{eff}^3$.

These results confirm that, for a constant $M_A$, the cavity always inflates when $\mathbf{B_\text{d}}$ and $\mathbf{B_\text{IMF}}$ are anti parallel and show that this inflation is more pronounced for low $M_A$. For asymptotically high $M_A$, the cavity size does not depend on the relative orientation between $\mathbf{B_\text{d}}$ and $\mathbf{B_\text{IMF}}$, as this corresponds to $B_\text{IMF} \to 0$. The simulation results are in good qualitative agreement with the analytical model. The error bars account for oscillations in the magnetopause position associated with the instabilities identified above, as well as for the fact that the cavity size is not perfectly circular. In Fig.~\ref{fig:2Dcavsize}, we can also observe that the region where shock formation is possible according to the criterion established in Sections~\ref{sec:constantrhoi_varyingL} and \ref{sec:constantL_varyingrhoi} (identified in grey) is very restrictive on the flow $M_A$ for small cavity sizes. However, the simulations with $M_A = 1.5$ (shown in Fig.~\ref{fig:bimforientation}) lie on this region (or very close to it) and these indeed show the formation of a shocked region in front of the cavity. Higher $M_A$ flows (\textit{e.g.} those represented in Figs.~\ref{fig:flowMA_zdipole} b) and c)) are far from this region and do not produce a shock. Finally, we note that this model does not directly account for the presence of a shocked region. For the cases with a clear shock formed ahead of the density cavity, this could be included by computing the downstream plasma density, temperature and magnetic field (using MHD conservation laws) and correcting the energy density balance in Eq.~(\ref{eq:Ebalance}). However, for most of the scenarios studied here, this would not be valid due to the non-stationarity of the shock front and the intrinsic kinetic character of the problem. Moreover, as shown in Figs.~\ref{fig:bdipolestrength} and \ref{fig:flowMA_zdipole}, we observe that the magnetopause position is not significantly altered in the presence of a shocked region and hence the model presented here gives a good qualitative description of the density cavity inflation/deflation depending on the relative orientation between $\mathbf{B_\text{d}}$ and $\mathbf{B_\text{IMF}}$.

\subsection{Role of field-aligned dynamics on magnetopause structure}
\label{sec:inplane}

The previous sections describe the plasma dynamics close to the magnetopause from a fundamental perspective. In all the simulations described above, $\mathbf{B_\text{IMF}}$ and $\mathbf{B_\text{d}}$ point in/out of the simulation plane, and thus they do not include important effects like field-aligned particle dynamics or the dipolar field curvature. In this section, we qualitatively describe the importance of these effects. The initial setup of the simulations presented in this section is similar to the one shown in Fig.~\ref{fig:setup_Bz}, with the obstacle dipolar and plasma internal magnetic fields lying in the simulation plane.

%\begin{figure}
%\includegraphics[height=6.8cm]{setup_BxBy.pdf}
%\caption{\label{fig:setup_BxBy} Initial setup of the 2D simulations with in-plane magnetic fields. The plasma is continuously injected from the left boundary and flows along the $x$ direction. Part of the plasma is initialized inside the box (indicated in grey), to reduce the computation effort. The colour map indicates the dipole field strength normalized to its maximum value, determined by a cutoff introduced in the dipolar field. The magnetic field line direction is indicated with arrows.}
%\end{figure}

\begin{figure*}
\includegraphics[height=7.1cm]{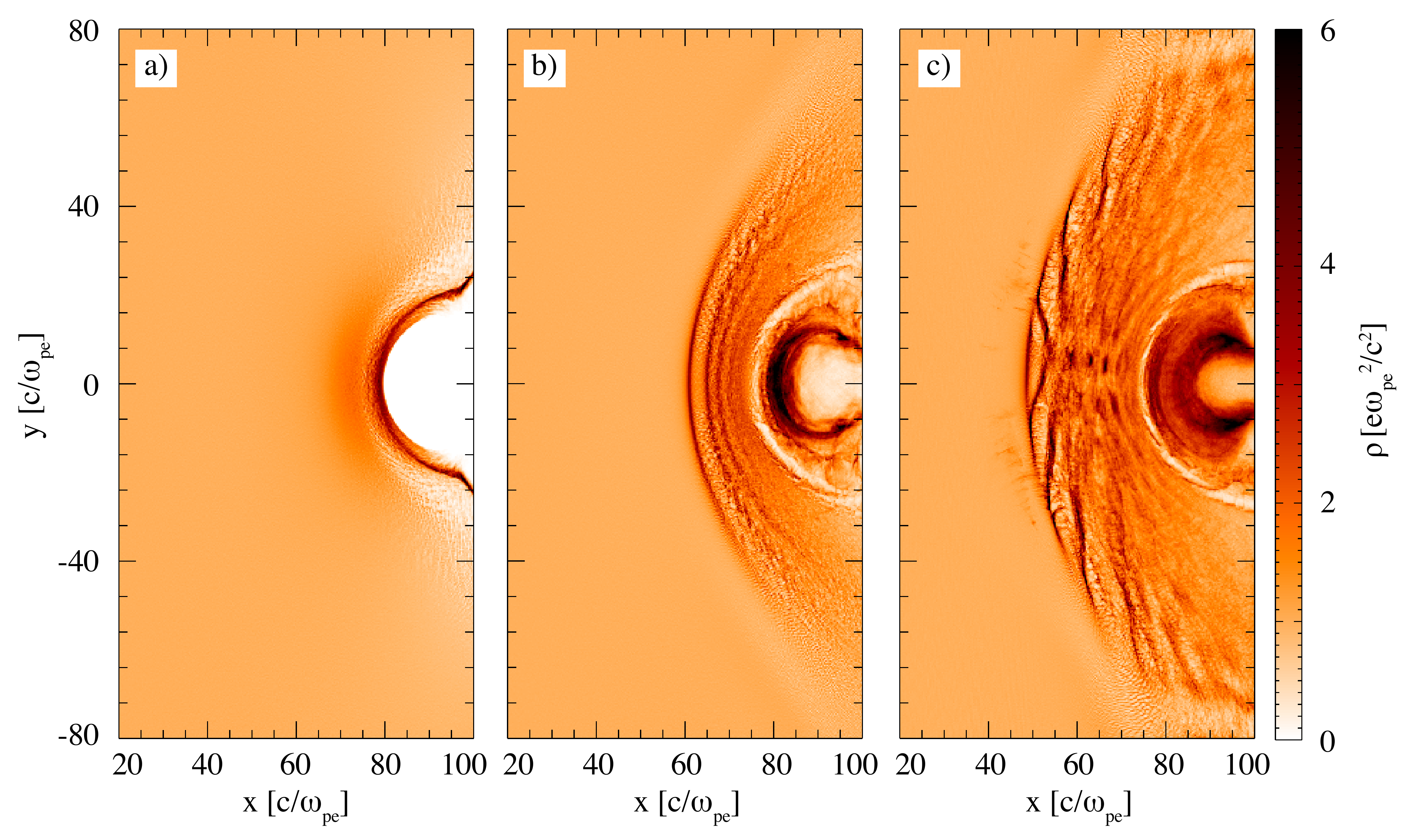}
\caption{\label{fig:inplane_par_timeevol} Time evolution of a plasma/dipolar magnetic field interaction with in-plane, parallel $\mathbf{B_\text{IMF}}$ and $\mathbf{B_\text{d}}$. A plasma flow with $v_0=0.1c$ and $M_A = 1.5$ is collided with a dipolar magnetic field that holds the plasma ram pressure at a distance $L_0 = 2 d_i$ according to the macroscopic pressure balance. The three panels represent the simulations times (a) $t \omega_{pe}=400$, (b) $t \omega_{pe}=800$ and (c) $t \omega_{pe}=1200$. Particle trapping in the in-plane field lines enhances the density pile-up close to the cavity and increases the effective obstacle size seen by the incoming plasma flow.}
\end{figure*}

\begin{figure*}
\includegraphics[height=7.1cm]{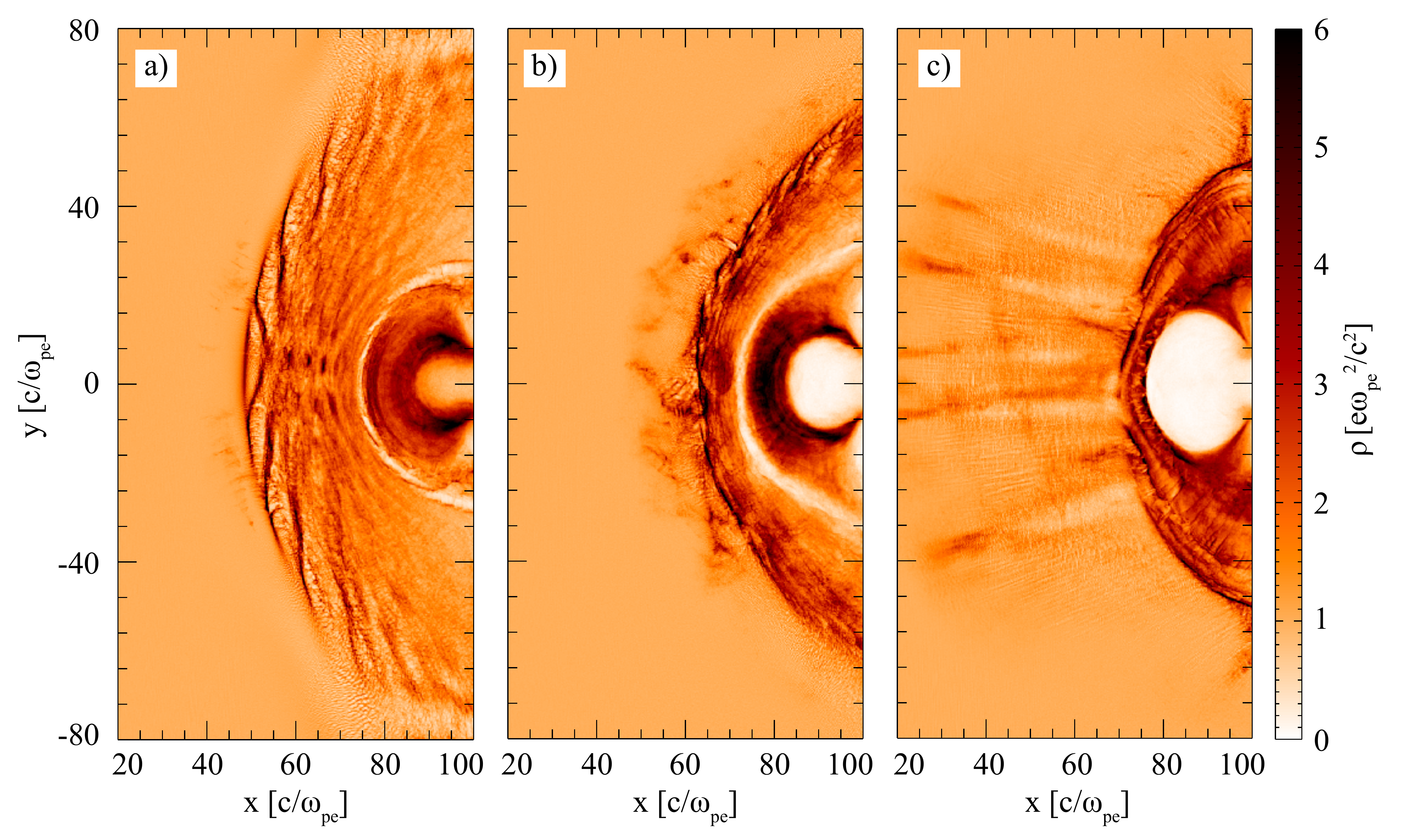}
\caption{\label{fig:flowMA_ydipole_par} Particle trapping increases effective cavity size. Three plasma flows with (a) $M_A = 1.5$, (b) $M_A = 3$ and (c) $M_A = 15$ interact with a magnetic dipole that standoff the plasma at a distance $L_0 = 2 d_i$. The in-plane fields $\mathbf{B_\text{IMF}}$ and $\mathbf{B_\text{d}}$ are parallel on the magnetosphere dayside.}
\end{figure*}

In Fig.~\ref{fig:inplane_par_timeevol}, we observe the time evolution of the interaction between a flow with $v_0=0.1c$ and $M_A = 1.5$ and a small-scale magnetic obstacle with in-plane $\mathbf{B_\text{IMF}}$ and $\mathbf{B_\text{d}}$. These fields are, in this case, parallel on the dayside of the magnetosphere (\textit{i.e.} on the direction from where the plasma collides with the magnetic obstacle). We can observe that, when the plasma is decelerated by the magnetic field ramp (see Fig.~\ref{fig:inplane_par_timeevol} a)), some particles are trapped in the in-plane field lines and recirculate in the inner region of the magnetosphere, contributing to the enhanced density pile-up observed in Figs.~\ref{fig:inplane_par_timeevol} b) and c). This trapping increases the effective size of the obstacle seen by the fresh impinging plasma and relaxes the condition on the maximum $M_A$ for shock formation. In Fig.~\ref{fig:flowMA_ydipole_par}, we show 2D simulation results of plasma flows with (a) $M_A = 1.5$, (b) $M_A = 3$ and (c) $M_A = 15$, all with parallel, in-plane $\mathbf{B_\text{IMF}}$ and $\mathbf{B_\text{d}}$. Although the density cavity is, in all the cases, smaller than the ion gyroradius, a compressed plasma region is observed for all considered $M_A$ flows. However, for higher $M_A$ flows, the compression is sufficient to reflect particles on the shock front, leading to modulations and instabilities on the transition between the unperturbed and shocked plasma regions (see Figs.~\ref{fig:flowMA_ydipole_par} b) and c)).

\begin{figure*}
\includegraphics[height=7.1cm]{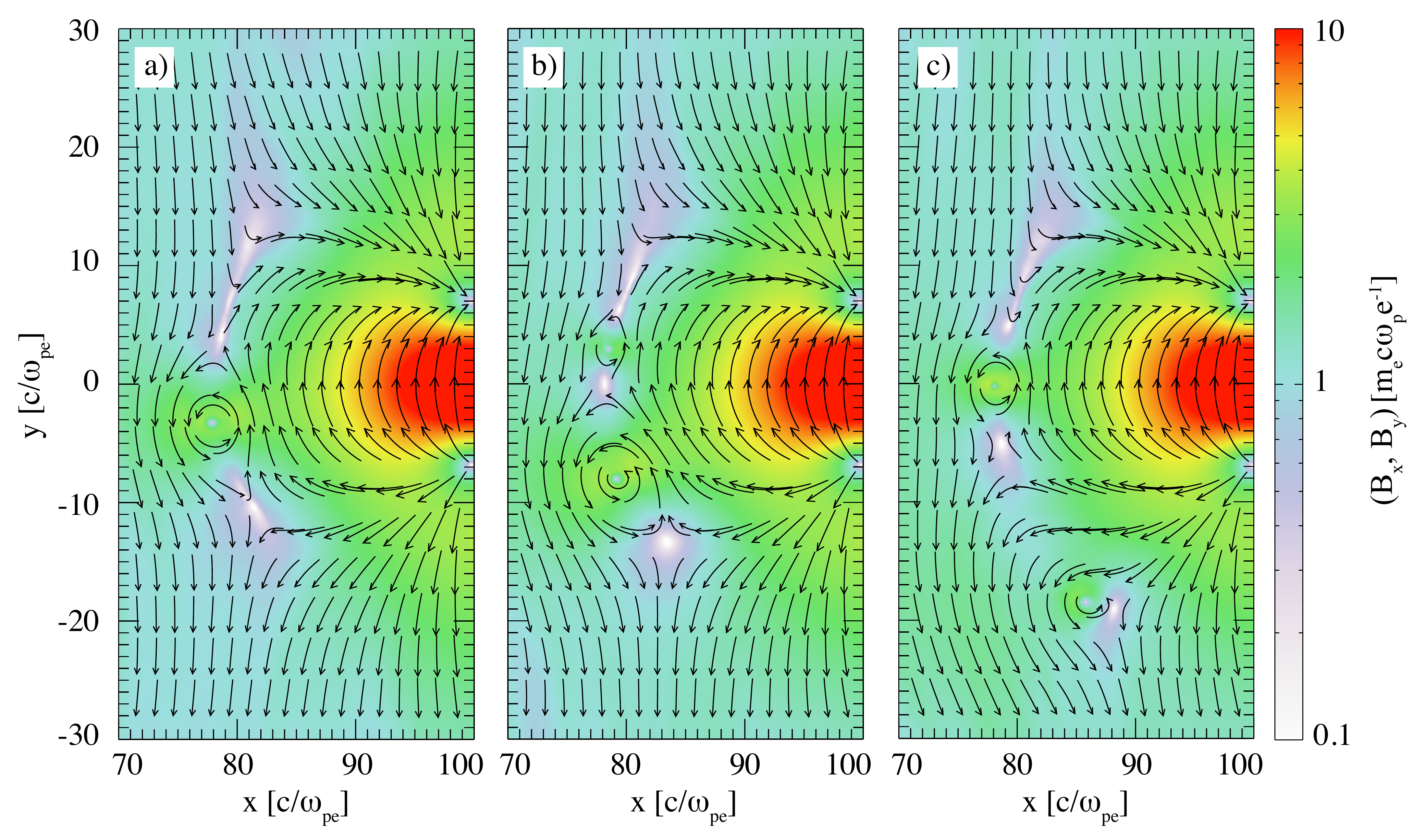}
\caption{\label{fig:inplane_antipar_B_timeevol} Time evolution of plasmoid formation and outflow in a plasma/dipolar magnetic field interaction with in-plane, anti parallel $\mathbf{B_\text{IMF}}$ and $\mathbf{B_\text{d}}$. A plasma flow with $v_0=0.1c$ and $M_A = 1.5$ is collided with a dipolar magnetic field that holds the plasma ram pressure at a distance $L_0 = 2 d_i$ according to the macroscopic pressure balance. The three panels show the total in-plane magnetic field magnitude (in colours) and direction (arrows) at the times (a) $t \omega_{pe}=1200$, (b) $t \omega_{pe}=1300$ and (c) $t \omega_{pe}=1400$. The reconnection product plasmoids move northwards/southwards (no preferential direction is observed) and are dragged to the region above the poles.}
\end{figure*}

\begin{figure*}
\includegraphics[height=7.1cm]{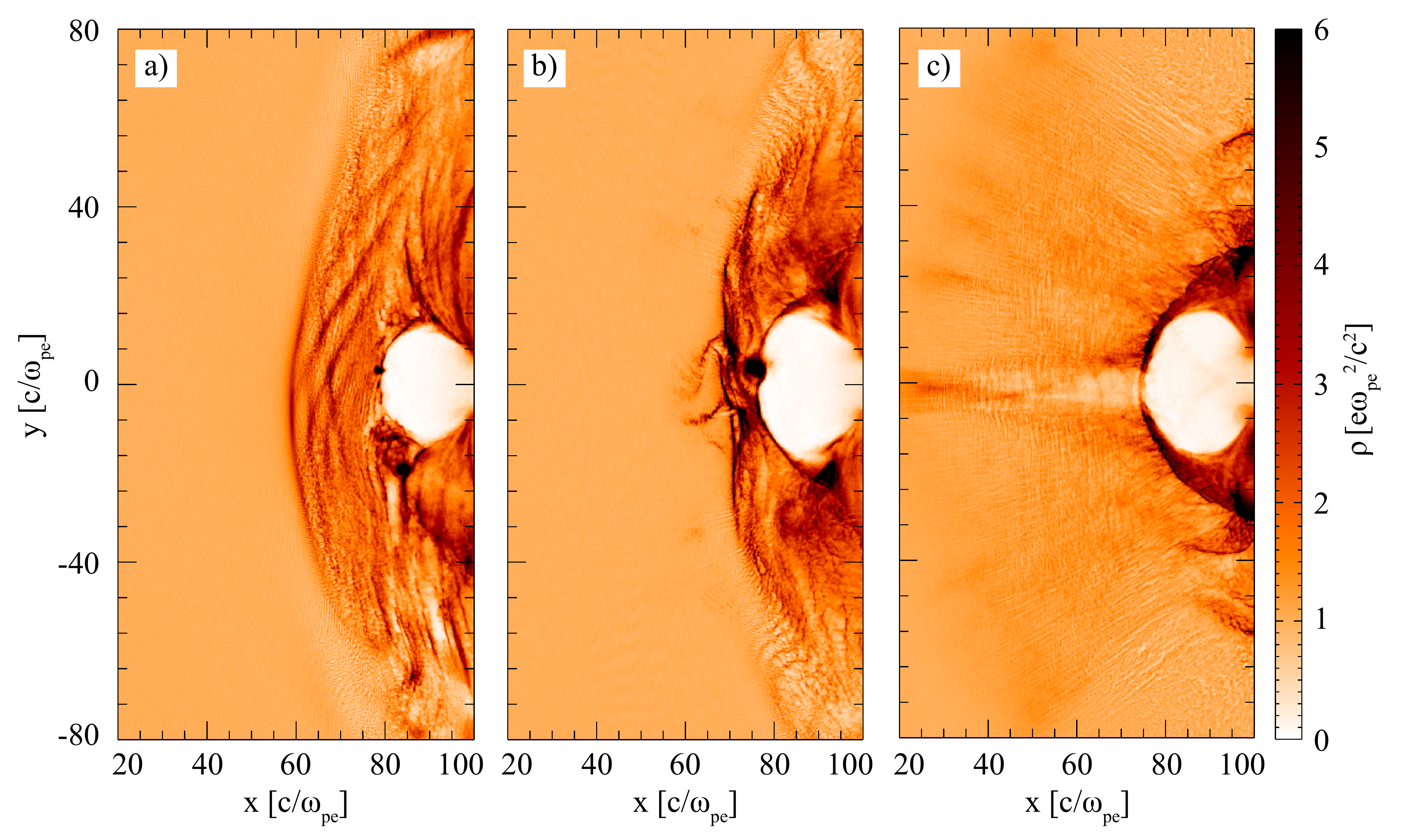}
\caption{\label{fig:flowMA_ydipole_antipar} Reconnection at the magnetopause prevents particle trapping. Three plasma flows with (a) $M_A = 1.5$, (b) $M_A = 3$ and (c) $M_A = 10$ interact with a magnetic dipole that standoff the plasma at a distance $L_0 = 2 d_i$. The in-plane fields $\mathbf{B_\text{IMF}}$ and $\mathbf{B_\text{d}}$ are anti parallel on the magnetosphere dayside and reconnection dominates the magnetopause dynamics.}
\end{figure*}

When $\mathbf{B_\text{IMF}}$ is anti parallel to $\mathbf{B_\text{d}}$, magnetic reconnection~\cite{Parker:1979, PriestForbes:2010, Yamada:2010} can occur on the magnetosphere dayside and the magnetopause dynamics changes dramatically from the parallel case described above. In Fig.~\ref{fig:inplane_antipar_B_timeevol}, we show the time evolution of the total in-plane magnetic field magnitude and direction (colour and arrow codes, respectively) for a 2D simulation of a plasma flow with $v_0=0.1c$ and $M_A = 1.5$. The orientation of $\mathbf{B_\text{IMF}}$ is inverted from those simulations in Figs.~\ref{fig:inplane_par_timeevol} and \ref{fig:flowMA_ydipole_par}. As the plasma compresses the dipolar field, the magnetic field lines are reconnected at the magnetopause and plasmoids emerge from the reconnected field lines~\cite{Markidis:2012, Loureiro:2007, Samtaney:2009}. These plasmoids are then dragged away from the magnetic null point. No preferential direction is observed for the dragging of the plasmoids. This is a mechanism of outflow for the reconnected field lines. In this case, no particle trapping is observed and the shock formation criterion defined in Sections~\ref{sec:constantrhoi_varyingL} and \ref{sec:constantL_varyingrhoi} remains the same. This is illustrated in Fig.~\ref{fig:flowMA_ydipole_antipar}, where 2D simulation results for different flow $M_A$ (and corresponding ion gyroradius $\rho_i = M_A d_i$) are compared. Like in Fig.~\ref{fig:flowMA_zdipole}, these simulations illustrate cases where (a) $L_0 > \rho_i$, (b) $L_0 \gtrsim \rho_i$ and (c) $L_0 < \rho_i$. We observe the formation of a clear shock for $L_0 > \rho_i$, the oscillatory dynamics described in Section~\ref{sec:constantrhoi_varyingL} for $L_0 \gtrsim \rho_i$ and the absence of a compressed plasma region for $L_0 < \rho_i$. In addition, in this plane we observe that the formation of plasmoids on the magnetopause can contribute to the breakdown of the oscillatory structure ahead of the magnetic obstacle, as the enhanced density pile-up deforms the plasma as it is formed and moves away from the magnetic null point (see Fig.~\ref{fig:flowMA_ydipole_antipar} b)).

The simulation results shown in Figs.~\ref{fig:inplane_par_timeevol} - \ref{fig:flowMA_ydipole_antipar} suggest that different microscopical, collective effects can play an important role in the magnetopause dynamics when the fields $\mathbf{B_\text{IMF}}$ and $\mathbf{B_\text{d}}$ are in-plane. Although these were not included in the analytical model for the cavity size described in Section~\ref{sec:IMForientation}, it is possible to make a qualitative analysis of the variation of the effective obstacle size as a function of the flow $M_A$. Even though the particle trapping results in the increase of the effective obstacle size observed for parallel $\mathbf{B_\text{IMF}}$ and $\mathbf{B_\text{d}}$, we can still observe an increase in the cavity size for higher $M_A$ in Fig.~\ref{fig:flowMA_ydipole_par}, in agreement with the idealized behaviour depicted in the physical picture of Fig.~\ref{fig:physpic}. Additionally, when reconnection is possible (\textit{i.e.} when $\mathbf{B_\text{IMF}}$ and $\mathbf{B_\text{d}}$ are anti parallel), the energy stored in the magnetic field is transferred to the particles in the form of kinetic energy parallel to the initial magnetic field. Since this happens at the magnetopause, a phenomenological term describing the energy density acquired due to reconnecting field lines $\varepsilon_\text{rec}$ can be added to the energy balance in Eq.~(\ref{eq:Ebalance}), giving
\begin{equation}
\label{eq:corEbalance}
\varepsilon_\text{rec} + n_0 \phi = n_0 m_i v_0^2 + \frac{B_\text{IMF}^2}{8\pi} \text{ .}
\end{equation}
The final outflow velocity is, by an energy density conservation argument, larger than $v_0$ by a factor $2L_0/\Delta \gg 1$. Thus, the energy density $\varepsilon_\text{rec}$ is large compared to the expected potential energy required to reflect the particles and a weaker dependance of the cavity size on the flow $M_A$ is expected. In fact, since the reconnection at the magnetopause may not be symmetric, the released energy density can, in general, be a complex function of the potential, \textit{i.e.} $\varepsilon_\text{rec} = \varepsilon_\text{rec} (\phi_0)$.

\subsection{Importance of 3D interplay in magnetopause and shock dynamics}
\label{sec:3Dinterplay}

Even though the qualitative analysis presented above is simplified, it gives important insights about a general, complex 3D scenario. Understanding the interplay between the two planes analysed separately in Sections~\ref{sec:constantrhoi_varyingL}-\ref{sec:constantL_varyingrhoi} and \ref{sec:inplane} is critical to describe the general dynamics of the particles at the magnetopause and the formation of shocks. In this section, we present 3D simulation results and qualitatively compare them with the corresponding 2D simulations.
We consider a plasma flow with $v_0 = 0.2c$ and $M_A = 1.5$ colliding with a dipolar magnetic field that stops the plasma at a distance $L_0 = 2 d_i$. In the 3D simulations, the plasma injection scheme A is used and the boundary conditions are the same as described in Section~\ref{sec:constantrhoi_varyingL}. The simulation domain has dimensions $L_x \times L_y \times L_z = 150 \times 300 \times 400 \ c/\omega_{pe}$ and the grid resolution is 3 cells/$d_e$ in all the directions, with 8 simulation particles per cell per species. To qualitatively evaluate the importance of the full three dimensional interplay in the magnetopause and shock dynamics, we performed 2D simulations of the two main interaction planes with the same flow parameters. These planes are the central ($y=0$ and $z=0$) slices perpendicular and parallel to the dipole magnetic moment, which is oriented in the positive $z$ direction for the 3D simulations.

\begin{figure*}
\includegraphics[height=9.8cm]{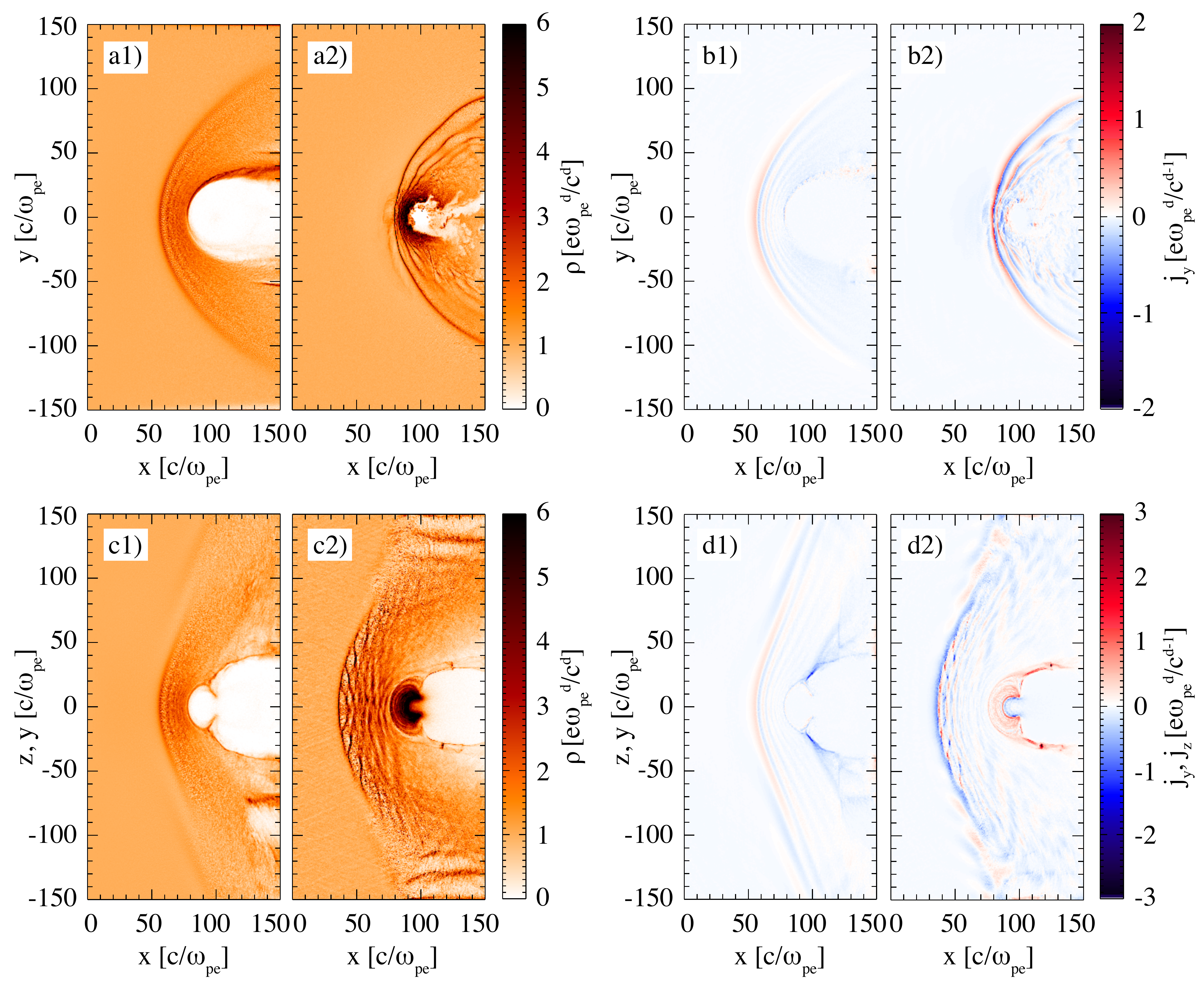}
\caption{\label{fig:3Dcomparison_par} Qualitative comparison between 2D and 3D simulations for parallel  $\mathbf{B_\text{IMF}}$ and $\mathbf{B_\text{d}}$. A plasma flow with $v_0=0.2c$ and $M_A = 1.5$ interacts with a magnetic dipole that reflects the plasma at a standoff distance of $L_0 = 2 d_i$. Panels a-d1 (on the left of each pair) show results of 3D simulations, whereas panels a-d2 show the corresponding 2D runs. Panels a and b (c and d) correspond to the central plane perpendicular (parallel) to the dipole magnetic moment. The time frames shown here correspond to the simulation time of $t \omega_{pe} = 1000$. The parameter $d$ in the proton and current density units corresponds to the number of dimensions of the simulation, \textit{i.e.} $d=3$ for panels a-d1 and $d=2$ for panels a-d2.}
\end{figure*}

For parallel $\mathbf{B_\text{IMF}}$ and $\mathbf{B_\text{d}}$ on the magnetopause dayside (see Fig.~\ref{fig:3Dcomparison_par}), we observe significantly different magnetopause dynamics. The particle trapping and high density pile-up close to the magnetopause registered in 2D simulations is not observed in the 3D cases (see panels a1,2 and c1,2 in Fig.~\ref{fig:3Dcomparison_par}), as the particles flow around the object in the transverse direction. The same features have a strong signature in the current perpendicular to the flow and the magnetic dipole moment (see, in particular, panel d2 in Fig.~\ref{fig:3Dcomparison_par}). The region in front of the obstacle is, however, qualitatively similar. A solitary-like perturbation (characteristic of such low $M_A$ interactions) is excited in this region in both 2D and 3D simulations, although it is clearer in 3D (see panels a-d1 in Fig.~\ref{fig:3Dcomparison_par}). In this case, the plasma compression is lower and the effective shock Mach number is smaller due to the stationarity of the perturbation (it propagates in the negative $x$ direction in 2D). The plasma discontinuity is slightly more elongated along the direction parallel to the dipole moment, which is consistent with the previous 2D qualitative comparison.

\begin{figure*}
\includegraphics[height=9.8cm]{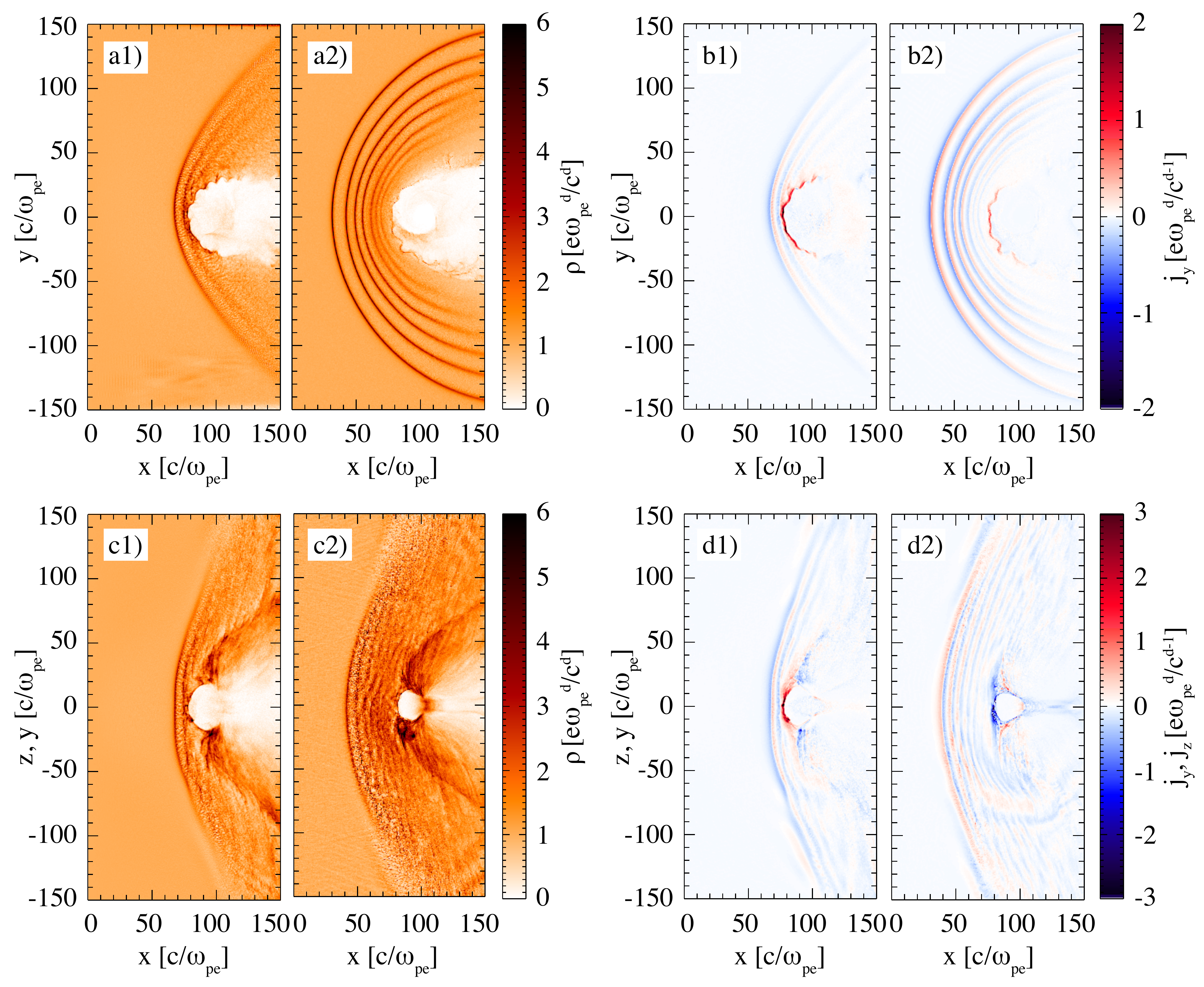}
\caption{\label{fig:3Dcomparison_antipar} Qualitative comparison between 2D and 3D simulations for anti parallel  $\mathbf{B_\text{IMF}}$ and $\mathbf{B_\text{d}}$. The plasma flow parameters and figure labels are the same as in Fig.~\ref{fig:3Dcomparison_par}.}
\end{figure*}

In Fig.~\ref{fig:3Dcomparison_antipar}, we show the same type of comparison between 2D and 3D simulations for a flow with $\mathbf{B_\text{IMF}}$ anti parallel to $\mathbf{B_\text{d}}$ on the magnetosphere dayside. In this case, the 2D and 3D simulations are also qualitatively similar. A current layer of width on the order of $d_e$ is present in all the planes (see panels b1,2 and d1,2 in Fig.~\ref{fig:3Dcomparison_antipar}). A solitary-like perturbation is excited in front of the magnetic obstacle. In this case, however, its shape is clearer in  the 2D simulations, as there is no particle pile-up and trapping in front of the object (see panels a1,2 and b1,2 in Fig.~\ref{fig:3Dcomparison_antipar}). The solitary perturbation propagates faster in 2D when the magnetic fields point out of the simulation plane (panels a2 and b2 in Fig.~\ref{fig:3Dcomparison_antipar}), due to the cavity inflation. For this reason, the time frames shown here correspond to earlier interaction times than those in the 3D simulations ($t_\text{2D} \omega_{pe} = 1100$, $t_\text{3D} \omega_{pe} = 1600$).

\begin{figure*}
\includegraphics[height=18.2cm]{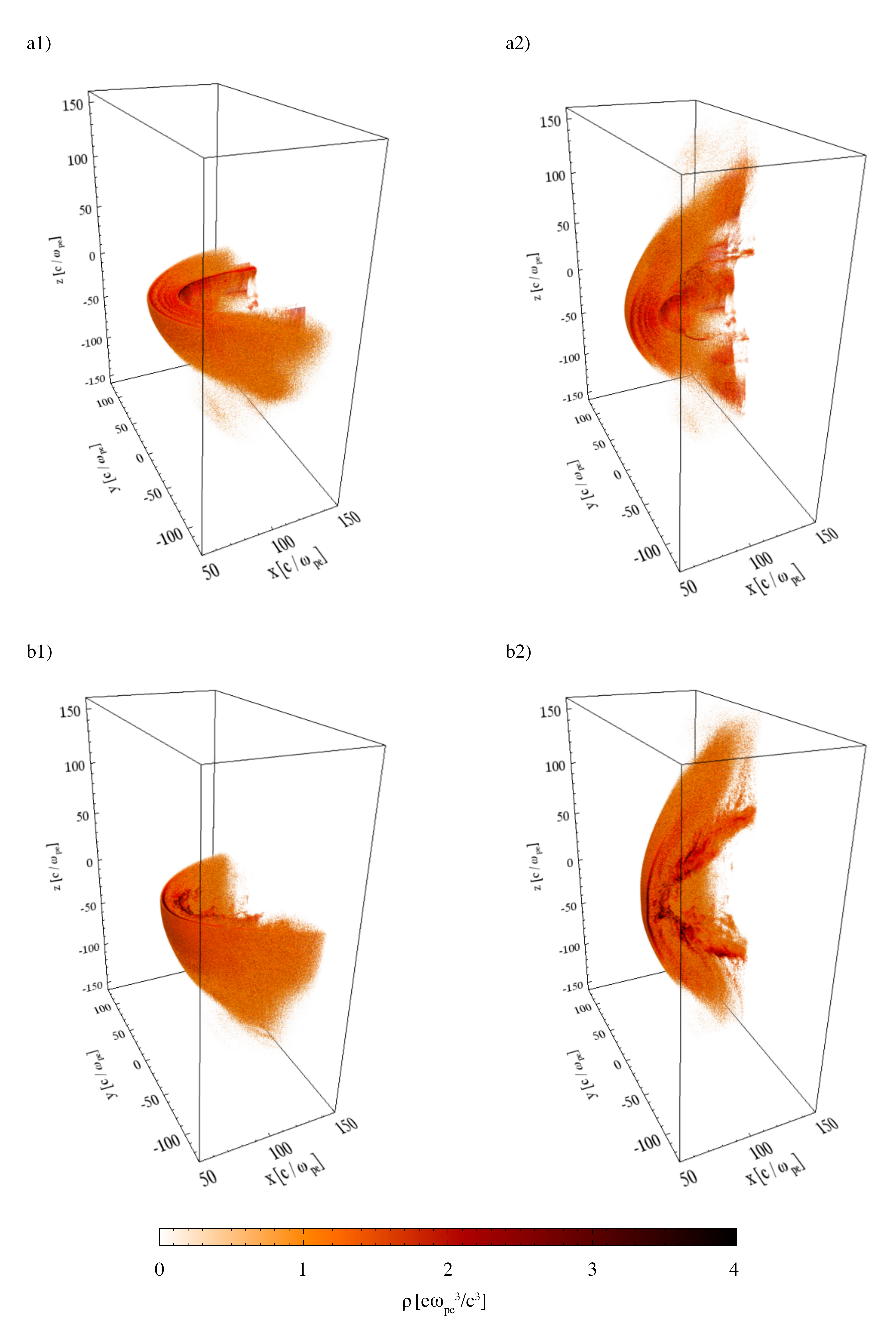}
\caption{\label{fig:3Dcomparison_3Dfigs} Three dimensional representation of the interaction between plasma flows with parallel and anti parallel $\mathbf{B_\text{IMF}}$ and $\mathbf{B_\text{d}}$ and a miniature obstacle. Panels a1-2 (b1-2) represent the plasma density volume rendering for the parallel (anti parallel) $\mathbf{B_\text{IMF}}$ and $\mathbf{B_\text{d}}$ case. Panels a1 and a2 (respectively b1 and b2) have a different clipping for visualisation purposes. The plasma flow parameters are the same as in Figs.~\ref{fig:3Dcomparison_par} and \ref{fig:3Dcomparison_antipar}.}
\end{figure*}

In general, a better qualitative agreement is found between the 3D and 2D simulations with in-plane magnetic fields. We find that the shock front is, in general elliptical. This effect is more pronounced when $\mathbf{B_\text{IMF}}$ is anti parallel to $\mathbf{B_\text{d}}$ due to the presence of higher density regions above and below the magnetic field poles (see panels c1,2 in Fig.~\ref{fig:3Dcomparison_antipar}). The higher density of these regions is enhanced due to the continuous formation and outflow of plasmoids in the magnetopause. A more graphical representation of the three dimensional interaction for the parallel and anti parallel cases is represented in Fig.~\ref{fig:3Dcomparison_3Dfigs}, illustrating these points.

\section{Laboratory parameters}
\label{sec:labparams}

\begin{figure}
\includegraphics[height=6.95cm]{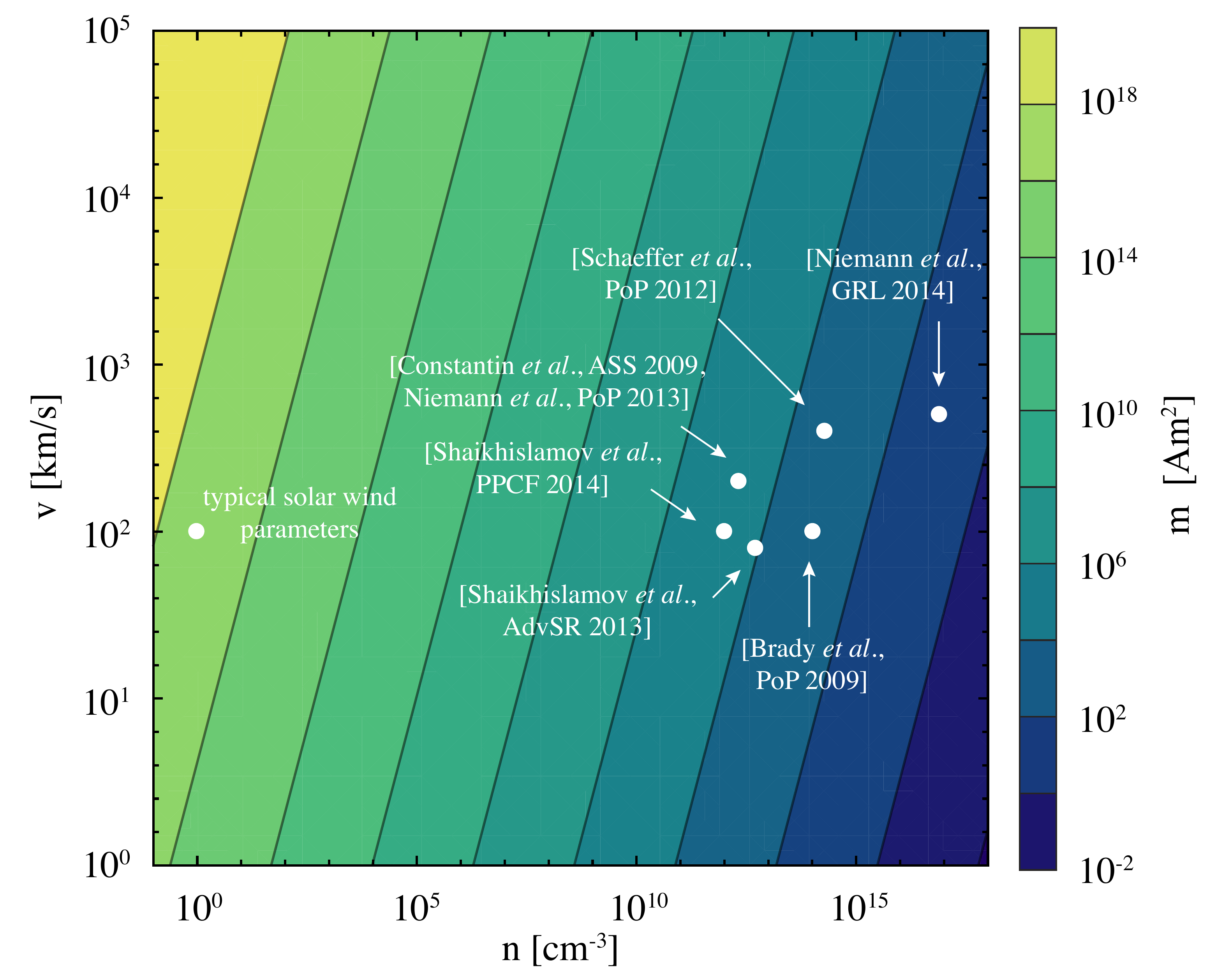}
\caption{\label{fig:labparams} Dipolar moment required to observe the formation of a collisionless shock with $M_A = 10$ in space and laboratory scenarios. The white dots represent typical flows for space and laboratory plasmas.}
\end{figure}

The analysis presented so far was focused on flows with parameters slightly different from those of space and laboratory plasmas, yet it is possible to infer about the microphysics of mini magnetospheres due to the system description in terms of the dimensionless quantities $L_\text{eff}/\rho_i$ and $M_A$. Considering now realistic parameters, we evaluate the possibility of generating shocks in laboratory and space scenarios. Taking the macroscopic pressure balance of Eq.~(\ref{eq:MHDpbalance}) and imposing a minimum cavity size of $L_\text{eff} =\rho_i$, we can find the required dipole magnetic moment $m$ to form a shock using a plasma flow with a velocity $v$, density $n$ and Mach number $M_A$. Using a realistic ion mass ratio of $m_i/m_e = 1836$, we obtain the results shown in Fig.~\ref{fig:labparams} for $M_A =10$. These results show that, for a constant density, higher flow velocities require higher dipolar moments to observe a shock, as we expect from the MHD pressure balance.

Interestingly, however, we can see that for a constant plasma velocity, an increase of the plasma density results in a decrease of $m$. This is a consequence of the fact that, by increasing $n$, we are not only increasing the plasma ram pressure, but we are also decreasing the plasma spatial scales, namely $d_i$. The competition between these two effects thus determines that higher plasma densities (for a constant flow velocity) are more favourable to observe shocks in the laboratory. Note that, for the plasma to have low a Mach number in high density conditions, it has to support high magnetic fields, of the order of $0.1 - 1$ T. This range of parameters is available, for instance, at the Large Plasma Device (LAPD)~\cite{Gekelman:1991, Niemann:2012}, or at the OMEGA laser facility~\cite{Waxer:2005}. In both these facilities, experimental studies on collisionless laboratory astrophysics were recently conducted, including the first experimental characterizations of collisionless magnetized shocks~\cite{Niemann:2014, Schaeffer:2016}.

In the plots presented in Fig.~\ref{fig:labparams}, we also show the typical parameters of the solar wind at 1 AU, as well as of laboratory plasmas produced in recent experiments~\cite{Brady:2009, Shaikhislamov:2009, Shaikhislamov:2013, Constantin:2009, Niemann:2013, Schaeffer:2012, Niemann:2014}. In the experiments where a magnetic obstacle was used~\cite{Brady:2009, Shaikhislamov:2009, Shaikhislamov:2013}, the formation of a density cavity was observed, although no shocks were registered. The magnets used in these experiments had a magnetic moment of about $60$ Am$^2$, \textit{i.e.} below (or very close to) the limit $m \simeq 10^2-10^3$ Am$^2$ estimated in Fig.~\ref{fig:labparams} for $M_A = 10$ (typical for the reported experimental parameters).

\section{Conclusions}
\label{sec:conclusions}

Recently available computational resources allow for an \textit{ab initio} approach to problems such as the formation of mini magnetospheres with sizes on the order of the plasma kinetic scales, a problem of relevance in space and astrophysical conditions, and also for ongoing experiments. In this work, we have shown that the critical obstacle size for the formation of shocks in mini magnetospheres is $L_\text{eff}/ \rho_i > 1$, resorting to massively parallel full PIC simulations. While the formation of a density cavity has been observed at sub-$\rho_i$ obstacles (in agreement with previous works~\cite{Deca:2014, Deca:2015, Bamford:2016}), we observe a distinct behaviour when $L/\rho_i>1$, with ions being able to recirculate in front of the obstacle, and ultimately enhancing the plasma compression in this region. We have demonstrated that the ratio $L_\text{eff}/ \rho_i$ can be controlled by both the dipolar moment and the ion Larmor radius (or equivalently the flow Alfv\`{e}nic Mach number). We have also explored the dynamics of the interaction in the transition between obstacles below and above the critical size. We have found that below, we observe a wake (consistent with Blanco-Cano \textit{et al.}~\cite{Blanco-Cano:2004}), at the transition, we find oscillatory dynamics, where there is periodic formation and dissipation of the shock, and above we see the formation of a well defined bow shock structure. Our results confirm that full PIC simulations accurately capture the interaction of plasma flows with magnetic obstacles, including the development of microinstabilities at the magnetopause (due to relative electron-ion streaming) and the microinstabilities triggered by the reflected plasma, which ultimately lead to the formation of the collisionless shocks.

We have also shown that the effective obstacle size may be strongly dependent on the relative orientation between the dipolar and plasma internal magnetic fields: anti parallel field configurations increase the effective size of the magnetic obstacle, whereas in parallel field configurations the effective size of the magnetic obstacle is decreased. This effect is particularly important in mini magnetospheres, as an inflation/deflation of few $d_i$ can be critical to allow/inhibit shock formation. The presence of an electron-scale current layer at the magnetopause was studied and used to model the kinetic cavity inflation/deflation. Our results suggest that, in general, space and laboratory small-scale systems may be highly mutable.

PIC simulations with in-plane magnetic fields showed additional features of the interaction. In particular, we found that, when $\mathbf{B_\text{IMF}}$ is anti parallel to $\mathbf{B_\text{d}}$ on the magnetosphere dayside, magnetic reconnection dominates the magnetopause interaction. The continuous formation of plasmoids at the magnetopause and consequent outflow to the poles of the magnetic field gives rise to an elongated compressed region in front of the object (when $L_\text{eff} > \rho_i$). We confirmed this result by comparing 2D and 3D simulations and we also showed that the shock dynamics can be investigated in 2D simulations. To understand the magnetopause dynamics, 3D simulations are necessary due to the intrinsic three dimensionality of the problem. However, 2D simulations can be used to give important insight about some of the features of the magnetopause (\textit{e.g.} the presence of the thin current layer and the reflecting electric field).

Finally, the possibility of generating collisionless shocks in laboratory and space scenarios has also been investigated, using the criterion for shock formation determined using 2D and 3D PIC simulations. We have shown that the required magnetic dipole moment to observe a shock in recent experiments is about one order of magnitude above the one considered. However, we expect that the collisionless shocks and the transition between the different interaction regimes studied in this work can be experimentally observed with recently available, highly magnetized plasma flows.

\begin{acknowledgments}
This work was supported by the European Research Council (InPairs ERC-2015-AdG 695088). Simulations were performed at the ACCELERATES cluster (Lisbon, Portugal).
\end{acknowledgments}

\bibliography{references}

\begin{thebibliography}{44}%
\makeatletter
\providecommand \@ifxundefined [1]{%
 \@ifx{#1\undefined}
}%
\providecommand \@ifnum [1]{%
 \ifnum #1\expandafter \@firstoftwo
 \else \expandafter \@secondoftwo
 \fi
}%
\providecommand \@ifx [1]{%
 \ifx #1\expandafter \@firstoftwo
 \else \expandafter \@secondoftwo
 \fi
}%
\providecommand \natexlab [1]{#1}%
\providecommand \enquote  [1]{``#1''}%
\providecommand \bibnamefont  [1]{#1}%
\providecommand \bibfnamefont [1]{#1}%
\providecommand \citenamefont [1]{#1}%
\providecommand \href@noop [0]{\@secondoftwo}%
\providecommand \href [0]{\begingroup \@sanitize@url \@href}%
\providecommand \@href[1]{\@@startlink{#1}\@@href}%
\providecommand \@@href[1]{\endgroup#1\@@endlink}%
\providecommand \@sanitize@url [0]{\catcode `\\12\catcode `\$12\catcode
  `\&12\catcode `\#12\catcode `\^12\catcode `\_12\catcode `\%12\relax}%
\providecommand \@@startlink[1]{}%
\providecommand \@@endlink[0]{}%
\providecommand \url  [0]{\begingroup\@sanitize@url \@url }%
\providecommand \@url [1]{\endgroup\@href {#1}{\urlprefix }}%
\providecommand \urlprefix  [0]{URL }%
\providecommand \Eprint [0]{\href }%
\providecommand \doibase [0]{http://dx.doi.org/}%
\providecommand \selectlanguage [0]{\@gobble}%
\providecommand \bibinfo  [0]{\@secondoftwo}%
\providecommand \bibfield  [0]{\@secondoftwo}%
\providecommand \translation [1]{[#1]}%
\providecommand \BibitemOpen [0]{}%
\providecommand \bibitemStop [0]{}%
\providecommand \bibitemNoStop [0]{.\EOS\space}%
\providecommand \EOS [0]{\spacefactor3000\relax}%
\providecommand \BibitemShut  [1]{\csname bibitem#1\endcsname}%
\let\auto@bib@innerbib\@empty
%</preamble>
\bibitem [{\citenamefont {Russel}(1991)}]{Russel:1991}%
  \BibitemOpen
  \bibfield  {author} {\bibinfo {author} {\bibfnamefont {C.~T.}\ \bibnamefont
  {Russel}},\ }\bibfield  {title} {\enquote {\bibinfo {title} {The
  magnetosphere},}\ }\href
  {http://dx.doi.org/10.1146/annurev.ea.19.050191.001125} {\bibfield  {journal}
  {\bibinfo  {journal} {Annu.\ Rev.\ Earth\ Planet.\ Sci.}\ }\textbf {\bibinfo
  {volume} {19}},\ \bibinfo {pages} {169--182} (\bibinfo {year}
  {1991})}\BibitemShut {NoStop}%
\bibitem [{\citenamefont {Colburn}\ \emph {et~al.}(1967)\citenamefont
  {Colburn}, \citenamefont {Currie}, \citenamefont {Mihalov},\ and\
  \citenamefont {Sonett}}]{Colburn:1967}%
  \BibitemOpen
  \bibfield  {author} {\bibinfo {author} {\bibfnamefont {D.~S.}\ \bibnamefont
  {Colburn}}, \bibinfo {author} {\bibfnamefont {R.~G.}\ \bibnamefont {Currie}},
  \bibinfo {author} {\bibfnamefont {J.~D.}\ \bibnamefont {Mihalov}}, \ and\
  \bibinfo {author} {\bibfnamefont {C.~P.}\ \bibnamefont {Sonett}},\ }\bibfield
   {title} {\enquote {\bibinfo {title} {Diamagnetic solar-wind cavity
  discovered behind moon},}\ }\href
  {http://dx.doi.org/10.1126/science.158.3804.1040} {\bibfield  {journal}
  {\bibinfo  {journal} {Science}\ }\textbf {\bibinfo {volume} {158}},\ \bibinfo
  {pages} {1040--1042} (\bibinfo {year} {1967})}\BibitemShut {NoStop}%
\bibitem [{\citenamefont {Lin}\ \emph {et~al.}(1998)\citenamefont {Lin},
  \citenamefont {Mitchell}, \citenamefont {Curtis}, \citenamefont {Anderson},
  \citenamefont {Carlson}, \citenamefont {McFadden}, \citenamefont {{n}a},
  \citenamefont {Hood},\ and\ \citenamefont {Binder}}]{Lin:1998}%
  \BibitemOpen
  \bibfield  {author} {\bibinfo {author} {\bibfnamefont {R.~P.}\ \bibnamefont
  {Lin}}, \bibinfo {author} {\bibfnamefont {D.~L.}\ \bibnamefont {Mitchell}},
  \bibinfo {author} {\bibfnamefont {D.~W.}\ \bibnamefont {Curtis}}, \bibinfo
  {author} {\bibfnamefont {K.~A.}\ \bibnamefont {Anderson}}, \bibinfo {author}
  {\bibfnamefont {C.~W.}\ \bibnamefont {Carlson}}, \bibinfo {author}
  {\bibfnamefont {J.}~\bibnamefont {McFadden}}, \bibinfo {author}
  {\bibfnamefont {M.~H.~A.}\ \bibnamefont {{n}a}}, \bibinfo {author}
  {\bibfnamefont {L.~L.}\ \bibnamefont {Hood}}, \ and\ \bibinfo {author}
  {\bibfnamefont {A.}~\bibnamefont {Binder}},\ }\bibfield  {title} {\enquote
  {\bibinfo {title} {Lunar surface magnetic fields and their interaction with
  the solar wind: Results from lunar prospector},}\ }\href
  {http://dx.doi.org/10.1126/science.281.5382.1480} {\bibfield  {journal}
  {\bibinfo  {journal} {Science}\ }\textbf {\bibinfo {volume} {281}},\ \bibinfo
  {pages} {1480--1484} (\bibinfo {year} {1998})}\BibitemShut {NoStop}%
\bibitem [{\citenamefont {Dwyer}, \citenamefont {Stevenson},\ and\
  \citenamefont {Nimmo}(2011)}]{Dwyer:2011}%
  \BibitemOpen
  \bibfield  {author} {\bibinfo {author} {\bibfnamefont {C.~A.}\ \bibnamefont
  {Dwyer}}, \bibinfo {author} {\bibfnamefont {D.~J.}\ \bibnamefont
  {Stevenson}}, \ and\ \bibinfo {author} {\bibfnamefont {F.}~\bibnamefont
  {Nimmo}},\ }\bibfield  {title} {\enquote {\bibinfo {title} {A long-lived
  lunar dynamo driven by continuous mechanical stirring},}\ }\href
  {http://dx.doi.org/10.1038/nature10564} {\bibfield  {journal} {\bibinfo
  {journal} {Nature}\ }\textbf {\bibinfo {volume} {479}},\ \bibinfo {pages}
  {212--214} (\bibinfo {year} {2011})}\BibitemShut {NoStop}%
\bibitem [{\citenamefont {Lillis}\ \emph {et~al.}(2013)\citenamefont {Lillis},
  \citenamefont {Robbins}, \citenamefont {Manga}, \citenamefont {Halekas},\
  and\ \citenamefont {Frey}}]{Lillis:2013}%
  \BibitemOpen
  \bibfield  {author} {\bibinfo {author} {\bibfnamefont {R.~J.}\ \bibnamefont
  {Lillis}}, \bibinfo {author} {\bibfnamefont {S.}~\bibnamefont {Robbins}},
  \bibinfo {author} {\bibfnamefont {M.}~\bibnamefont {Manga}}, \bibinfo
  {author} {\bibfnamefont {J.~S.}\ \bibnamefont {Halekas}}, \ and\ \bibinfo
  {author} {\bibfnamefont {H.~V.}\ \bibnamefont {Frey}},\ }\bibfield  {title}
  {\enquote {\bibinfo {title} {Time history of the martian dynamo from crater
  magnetic field analysis},}\ }\href {http://dx.doi.org/10.1002/jgre.20105}
  {\bibfield  {journal} {\bibinfo  {journal} {J.\ Geophys.\ Res.\ Planets}\
  }\textbf {\bibinfo {volume} {118}},\ \bibinfo {pages} {1488--1511} (\bibinfo
  {year} {2013})}\BibitemShut {NoStop}%
\bibitem [{\citenamefont {Russell}\ \emph {et~al.}(1984)\citenamefont
  {Russell}, \citenamefont {Phillips}, \citenamefont {Arghavani}, \citenamefont
  {Mihalov}, \citenamefont {Knudsen},\ and\ \citenamefont
  {Miller}}]{Russel:1984}%
  \BibitemOpen
  \bibfield  {author} {\bibinfo {author} {\bibfnamefont {C.~T.}\ \bibnamefont
  {Russell}}, \bibinfo {author} {\bibfnamefont {J.~L.}\ \bibnamefont
  {Phillips}}, \bibinfo {author} {\bibfnamefont {M.~R.}\ \bibnamefont
  {Arghavani}}, \bibinfo {author} {\bibfnamefont {J.~D.}\ \bibnamefont
  {Mihalov}}, \bibinfo {author} {\bibfnamefont {W.~C.}\ \bibnamefont
  {Knudsen}}, \ and\ \bibinfo {author} {\bibfnamefont {K.}~\bibnamefont
  {Miller}},\ }\bibfield  {title} {\enquote {\bibinfo {title} {A possible
  observation of a cometary bow shock},}\ }\href
  {http://dx.doi.org/10.1029/GL011i010p01022} {\bibfield  {journal} {\bibinfo
  {journal} {Geophys.\ Res.\ Lett.}\ }\textbf {\bibinfo {volume} {11}},\
  \bibinfo {pages} {1022--1025} (\bibinfo {year} {1984})}\BibitemShut {NoStop}%
\bibitem [{\citenamefont {Adams}\ \emph {et~al.}(2005)\citenamefont {Adams},
  \citenamefont {Hathaway}, \citenamefont {Grugel}, \citenamefont {Watts},
  \citenamefont {Parnell}, \citenamefont {Gregory},\ and\ \citenamefont
  {Winglee}}]{Adams:2005}%
  \BibitemOpen
  \bibfield  {author} {\bibinfo {author} {\bibfnamefont {J.~H.}\ \bibnamefont
  {Adams}}, \bibinfo {author} {\bibfnamefont {D.~H.}\ \bibnamefont {Hathaway}},
  \bibinfo {author} {\bibfnamefont {R.~N.}\ \bibnamefont {Grugel}}, \bibinfo
  {author} {\bibfnamefont {J.~W.}\ \bibnamefont {Watts}}, \bibinfo {author}
  {\bibfnamefont {T.~A.}\ \bibnamefont {Parnell}}, \bibinfo {author}
  {\bibfnamefont {J.~C.}\ \bibnamefont {Gregory}}, \ and\ \bibinfo {author}
  {\bibfnamefont {R.~M.}\ \bibnamefont {Winglee}},\ }\bibfield  {title}
  {\enquote {\bibinfo {title} {Revolutionary concepts of radiation shielding
  for human exploration of space},}\ }\href@noop {} {\bibfield  {journal}
  {\bibinfo  {journal} {NASA/TM-2005-213688, M-1133}\ } (\bibinfo {year}
  {2005})}\BibitemShut {NoStop}%
\bibitem [{\citenamefont {Parker}(2006)}]{Parker:2006}%
  \BibitemOpen
  \bibfield  {author} {\bibinfo {author} {\bibfnamefont {E.~N.}\ \bibnamefont
  {Parker}},\ }\bibfield  {title} {\enquote {\bibinfo {title} {Shielding space
  travelers},}\ }\href {http://dx.doi.org/10.1038/scientificamerican0306-40}
  {\bibfield  {journal} {\bibinfo  {journal} {Sci.\ Am.}\ }\textbf {\bibinfo
  {volume} {294}},\ \bibinfo {pages} {40---47} (\bibinfo {year}
  {2006})}\BibitemShut {NoStop}%
\bibitem [{\citenamefont {Bamford}\ \emph {et~al.}(2014)\citenamefont
  {Bamford}, \citenamefont {Kellett}, \citenamefont {Bradford}, \citenamefont
  {Todd}, \citenamefont {Sr.}, \citenamefont {Stafford-Allen}, \citenamefont
  {Alves}, \citenamefont {Silva}, \citenamefont {Collingwood}, \citenamefont
  {Crawford},\ and\ \citenamefont {Bingham}}]{Bamford:2014}%
  \BibitemOpen
  \bibfield  {author} {\bibinfo {author} {\bibfnamefont {R.~A.}\ \bibnamefont
  {Bamford}}, \bibinfo {author} {\bibfnamefont {B.}~\bibnamefont {Kellett}},
  \bibinfo {author} {\bibfnamefont {J.}~\bibnamefont {Bradford}}, \bibinfo
  {author} {\bibfnamefont {T.~N.}\ \bibnamefont {Todd}}, \bibinfo {author}
  {\bibfnamefont {M.~G.~B.}\ \bibnamefont {Sr.}}, \bibinfo {author}
  {\bibfnamefont {R.}~\bibnamefont {Stafford-Allen}}, \bibinfo {author}
  {\bibfnamefont {E.~P.}\ \bibnamefont {Alves}}, \bibinfo {author}
  {\bibfnamefont {L.~O.}\ \bibnamefont {Silva}}, \bibinfo {author}
  {\bibfnamefont {C.}~\bibnamefont {Collingwood}}, \bibinfo {author}
  {\bibfnamefont {I.~A.}\ \bibnamefont {Crawford}}, \ and\ \bibinfo {author}
  {\bibfnamefont {R.}~\bibnamefont {Bingham}},\ }\bibfield  {title} {\enquote
  {\bibinfo {title} {An exploration of the effectiveness of artificial
  mini-magnetospheres as a potential solar storm shelter for long term human
  space missions},}\ }\href {http://dx.doi.org/10.1016/j.actaastro.2014.10.012}
  {\bibfield  {journal} {\bibinfo  {journal} {Acta\ Astronaut.}\ }\textbf
  {\bibinfo {volume} {105}},\ \bibinfo {pages} {385--394} (\bibinfo {year}
  {2014})}\BibitemShut {NoStop}%
\bibitem [{\citenamefont {Winglee}\ \emph {et~al.}(2000)\citenamefont
  {Winglee}, \citenamefont {Slough}, \citenamefont {Ziemba},\ and\
  \citenamefont {Goodson}}]{Winglee:2000}%
  \BibitemOpen
  \bibfield  {author} {\bibinfo {author} {\bibfnamefont {R.~M.}\ \bibnamefont
  {Winglee}}, \bibinfo {author} {\bibfnamefont {J.}~\bibnamefont {Slough}},
  \bibinfo {author} {\bibfnamefont {T.}~\bibnamefont {Ziemba}}, \ and\ \bibinfo
  {author} {\bibfnamefont {A.}~\bibnamefont {Goodson}},\ }\bibfield  {title}
  {\enquote {\bibinfo {title} {Mini-magnetospheric plasma propulsion: Tapping
  the energy of the solar wind for spacecraft propulsion},}\ }\href
  {http://dx.doi.org/10.1029/1999JA000334} {\bibfield  {journal} {\bibinfo
  {journal} {J.\ Geophys.\ Res.}\ }\textbf {\bibinfo {volume} {105}},\ \bibinfo
  {pages} {21067–--21077} (\bibinfo {year} {2000})}\BibitemShut {NoStop}%
\bibitem [{\citenamefont {Brady}\ \emph {et~al.}(2009)\citenamefont {Brady},
  \citenamefont {Ditmire}, \citenamefont {Horton}, \citenamefont {Mays},\ and\
  \citenamefont {Zakharov}}]{Brady:2009}%
  \BibitemOpen
  \bibfield  {author} {\bibinfo {author} {\bibfnamefont {P.}~\bibnamefont
  {Brady}}, \bibinfo {author} {\bibfnamefont {T.}~\bibnamefont {Ditmire}},
  \bibinfo {author} {\bibfnamefont {W.}~\bibnamefont {Horton}}, \bibinfo
  {author} {\bibfnamefont {M.~L.}\ \bibnamefont {Mays}}, \ and\ \bibinfo
  {author} {\bibfnamefont {Y.}~\bibnamefont {Zakharov}},\ }\bibfield  {title}
  {\enquote {\bibinfo {title} {Laboratory experiments simulating solar wind
  driven magnetospheres},}\ }\href {http://dx.doi.org/10.1063/1.3085786}
  {\bibfield  {journal} {\bibinfo  {journal} {Phys.\ Plasmas}\ }\textbf
  {\bibinfo {volume} {16}},\ \bibinfo {pages} {043112} (\bibinfo {year}
  {2009})}\BibitemShut {NoStop}%
\bibitem [{\citenamefont {Shaikhislamov}\ \emph {et~al.}(2009)\citenamefont
  {Shaikhislamov}, \citenamefont {Antonov}, \citenamefont {Zakharov},
  \citenamefont {Boyarintsev}, \citenamefont {Melekhov}, \citenamefont
  {Posukh},\ and\ \citenamefont {Ponomarenko}}]{Shaikhislamov:2009}%
  \BibitemOpen
  \bibfield  {author} {\bibinfo {author} {\bibfnamefont {I.~F.}\ \bibnamefont
  {Shaikhislamov}}, \bibinfo {author} {\bibfnamefont {V.~M.}\ \bibnamefont
  {Antonov}}, \bibinfo {author} {\bibfnamefont {Y.~P.}\ \bibnamefont
  {Zakharov}}, \bibinfo {author} {\bibfnamefont {E.~L.}\ \bibnamefont
  {Boyarintsev}}, \bibinfo {author} {\bibfnamefont {A.~V.}\ \bibnamefont
  {Melekhov}}, \bibinfo {author} {\bibfnamefont {V.~G.}\ \bibnamefont
  {Posukh}}, \ and\ \bibinfo {author} {\bibfnamefont {A.~G.}\ \bibnamefont
  {Ponomarenko}},\ }\bibfield  {title} {\enquote {\bibinfo {title} {Laboratory
  simulation of field aligned currents in an experiment on laser-produced
  plasma interacting with a magnetic dipole},}\ }\href
  {http://dx.doi.org/10.1088/0741-3335/51/10/105005} {\bibfield  {journal}
  {\bibinfo  {journal} {Plasma\ Phys.\ Controlled\ Fusion}\ }\textbf {\bibinfo
  {volume} {51}},\ \bibinfo {pages} {105005} (\bibinfo {year}
  {2009})}\BibitemShut {NoStop}%
\bibitem [{\citenamefont {Shaikhislamov}\ \emph {et~al.}(2013)\citenamefont
  {Shaikhislamov}, \citenamefont {Antonov}, \citenamefont {Zakharov},
  \citenamefont {Boyarintsev}, \citenamefont {Melekhov}, \citenamefont
  {Posukh},\ and\ \citenamefont {Ponomarenko}}]{Shaikhislamov:2013}%
  \BibitemOpen
  \bibfield  {author} {\bibinfo {author} {\bibfnamefont {I.~F.}\ \bibnamefont
  {Shaikhislamov}}, \bibinfo {author} {\bibfnamefont {V.~M.}\ \bibnamefont
  {Antonov}}, \bibinfo {author} {\bibfnamefont {Y.~P.}\ \bibnamefont
  {Zakharov}}, \bibinfo {author} {\bibfnamefont {E.~L.}\ \bibnamefont
  {Boyarintsev}}, \bibinfo {author} {\bibfnamefont {A.~V.}\ \bibnamefont
  {Melekhov}}, \bibinfo {author} {\bibfnamefont {V.~G.}\ \bibnamefont
  {Posukh}}, \ and\ \bibinfo {author} {\bibfnamefont {A.~G.}\ \bibnamefont
  {Ponomarenko}},\ }\bibfield  {title} {\enquote {\bibinfo {title}
  {Mini-magnetosphere: Laboratory experiment, physical model and hall mhd
  simulation},}\ }\href {http://dx.doi.org/10.1016/j.asr.2013.03.034}
  {\bibfield  {journal} {\bibinfo  {journal} {Adv.\ Space\ Res.}\ }\textbf
  {\bibinfo {volume} {52}},\ \bibinfo {pages} {422--436} (\bibinfo {year}
  {2013})}\BibitemShut {NoStop}%
\bibitem [{\citenamefont {Bamford}\ \emph {et~al.}(2012)\citenamefont
  {Bamford}, \citenamefont {Kellett}, \citenamefont {Bradford}, \citenamefont
  {Norberg}, \citenamefont {Thornton}, \citenamefont {Gibson}, \citenamefont
  {Crawford}, \citenamefont {Silva}, \citenamefont {Gargat\'e},\ and\
  \citenamefont {Bingham}}]{Bamford:2012}%
  \BibitemOpen
  \bibfield  {author} {\bibinfo {author} {\bibfnamefont {R.~A.}\ \bibnamefont
  {Bamford}}, \bibinfo {author} {\bibfnamefont {B.}~\bibnamefont {Kellett}},
  \bibinfo {author} {\bibfnamefont {W.~J.}\ \bibnamefont {Bradford}}, \bibinfo
  {author} {\bibfnamefont {C.}~\bibnamefont {Norberg}}, \bibinfo {author}
  {\bibfnamefont {A.}~\bibnamefont {Thornton}}, \bibinfo {author}
  {\bibfnamefont {K.~J.}\ \bibnamefont {Gibson}}, \bibinfo {author}
  {\bibfnamefont {I.~A.}\ \bibnamefont {Crawford}}, \bibinfo {author}
  {\bibfnamefont {L.~O.}\ \bibnamefont {Silva}}, \bibinfo {author}
  {\bibfnamefont {L.}~\bibnamefont {Gargat\'e}}, \ and\ \bibinfo {author}
  {\bibfnamefont {R.}~\bibnamefont {Bingham}},\ }\bibfield  {title} {\enquote
  {\bibinfo {title} {Minimagnetospheres above the lunar surface and the
  formation of lunar swirls},}\ }\href
  {http://dx.doi.org/10.1103/PhysRevLett.109.081101} {\bibfield  {journal}
  {\bibinfo  {journal} {Phys.\ Rev.\ Lett.}\ }\textbf {\bibinfo {volume}
  {109}},\ \bibinfo {pages} {081101} (\bibinfo {year} {2012})}\BibitemShut
  {NoStop}%
\bibitem [{\citenamefont {Gargat\'e}\ \emph {et~al.}(2008)\citenamefont
  {Gargat\'e}, \citenamefont {Bingham}, \citenamefont {Fonseca}, \citenamefont
  {Bamford}, \citenamefont {Thornton}, \citenamefont {Gibson}, \citenamefont
  {Bradford},\ and\ \citenamefont {Silva}}]{Gargate:2008}%
  \BibitemOpen
  \bibfield  {author} {\bibinfo {author} {\bibfnamefont {L.}~\bibnamefont
  {Gargat\'e}}, \bibinfo {author} {\bibfnamefont {R.}~\bibnamefont {Bingham}},
  \bibinfo {author} {\bibfnamefont {R.~A.}\ \bibnamefont {Fonseca}}, \bibinfo
  {author} {\bibfnamefont {R.~A.}\ \bibnamefont {Bamford}}, \bibinfo {author}
  {\bibfnamefont {A.}~\bibnamefont {Thornton}}, \bibinfo {author}
  {\bibfnamefont {K.}~\bibnamefont {Gibson}}, \bibinfo {author} {\bibfnamefont
  {J.}~\bibnamefont {Bradford}}, \ and\ \bibinfo {author} {\bibfnamefont
  {L.~O.}\ \bibnamefont {Silva}},\ }\bibfield  {title} {\enquote {\bibinfo
  {title} {Hybrid simulations of mini-magnetospheres in the laboratory},}\
  }\href {http://dx.doi.org/10.1088/0741-3335/50/7/074017} {\bibfield
  {journal} {\bibinfo  {journal} {Plasma\ Phys.\ Controlled\ Fusion}\ }\textbf
  {\bibinfo {volume} {50}},\ \bibinfo {pages} {074017} (\bibinfo {year}
  {2008})}\BibitemShut {NoStop}%
\bibitem [{\citenamefont {Kugland}\ \emph {et~al.}(2012)\citenamefont
  {Kugland}, \citenamefont {Ryutov}, \citenamefont {Chang}, \citenamefont
  {Drake}, \citenamefont {Fiksel}, \citenamefont {Froula}, \citenamefont
  {Glenzer}, \citenamefont {Gregori}, \citenamefont {Grosskopf}, \citenamefont
  {Koenig}, \citenamefont {Kuramitsu}, \citenamefont {Kuranz}, \citenamefont
  {Levy}, \citenamefont {Liang}, \citenamefont {Meinecke}, \citenamefont
  {Miniati}, \citenamefont {Morita}, \citenamefont {Pelka}, \citenamefont
  {Plechaty}, \citenamefont {Presura}, \citenamefont {Ravasio}, \citenamefont
  {Remington}, \citenamefont {Reville}, \citenamefont {Ross}, \citenamefont
  {Sakawa}, \citenamefont {Spitkovsky}, \citenamefont {Takabe},\ and\
  \citenamefont {Park}}]{Kugland:2012}%
  \BibitemOpen
  \bibfield  {author} {\bibinfo {author} {\bibfnamefont {N.~L.}\ \bibnamefont
  {Kugland}}, \bibinfo {author} {\bibfnamefont {D.~D.}\ \bibnamefont {Ryutov}},
  \bibinfo {author} {\bibfnamefont {P.-Y.}\ \bibnamefont {Chang}}, \bibinfo
  {author} {\bibfnamefont {R.~P.}\ \bibnamefont {Drake}}, \bibinfo {author}
  {\bibfnamefont {G.}~\bibnamefont {Fiksel}}, \bibinfo {author} {\bibfnamefont
  {D.~H.}\ \bibnamefont {Froula}}, \bibinfo {author} {\bibfnamefont {S.~H.}\
  \bibnamefont {Glenzer}}, \bibinfo {author} {\bibfnamefont {G.}~\bibnamefont
  {Gregori}}, \bibinfo {author} {\bibfnamefont {M.}~\bibnamefont {Grosskopf}},
  \bibinfo {author} {\bibfnamefont {M.}~\bibnamefont {Koenig}}, \bibinfo
  {author} {\bibfnamefont {Y.}~\bibnamefont {Kuramitsu}}, \bibinfo {author}
  {\bibfnamefont {C.}~\bibnamefont {Kuranz}}, \bibinfo {author} {\bibfnamefont
  {M.~C.}\ \bibnamefont {Levy}}, \bibinfo {author} {\bibfnamefont
  {E.}~\bibnamefont {Liang}}, \bibinfo {author} {\bibfnamefont
  {J.}~\bibnamefont {Meinecke}}, \bibinfo {author} {\bibfnamefont
  {F.}~\bibnamefont {Miniati}}, \bibinfo {author} {\bibfnamefont
  {T.}~\bibnamefont {Morita}}, \bibinfo {author} {\bibfnamefont
  {A.}~\bibnamefont {Pelka}}, \bibinfo {author} {\bibfnamefont
  {C.}~\bibnamefont {Plechaty}}, \bibinfo {author} {\bibfnamefont
  {R.}~\bibnamefont {Presura}}, \bibinfo {author} {\bibfnamefont
  {A.}~\bibnamefont {Ravasio}}, \bibinfo {author} {\bibfnamefont {B.~A.}\
  \bibnamefont {Remington}}, \bibinfo {author} {\bibfnamefont {B.}~\bibnamefont
  {Reville}}, \bibinfo {author} {\bibfnamefont {J.~S.}\ \bibnamefont {Ross}},
  \bibinfo {author} {\bibfnamefont {Y.}~\bibnamefont {Sakawa}}, \bibinfo
  {author} {\bibfnamefont {A.}~\bibnamefont {Spitkovsky}}, \bibinfo {author}
  {\bibfnamefont {H.}~\bibnamefont {Takabe}}, \ and\ \bibinfo {author}
  {\bibfnamefont {H.-S.}\ \bibnamefont {Park}},\ }\bibfield  {title} {\enquote
  {\bibinfo {title} {Self-organized electromagnetic field structures in
  laser-produced counter-streaming plasmas},}\ }\href
  {http://dx.doi.org/10.1038/nphys2434} {\bibfield  {journal} {\bibinfo
  {journal} {Nature\ Phys.}\ }\textbf {\bibinfo {volume} {8}},\ \bibinfo
  {pages} {809--812} (\bibinfo {year} {2012})}\BibitemShut {NoStop}%
\bibitem [{\citenamefont {Fox}\ \emph {et~al.}(2013)\citenamefont {Fox},
  \citenamefont {Fiksel}, \citenamefont {Bhattacharjee}, \citenamefont {Chang},
  \citenamefont {Germaschewski}, \citenamefont {Hu},\ and\ \citenamefont
  {Nilson}}]{Fox:2013}%
  \BibitemOpen
  \bibfield  {author} {\bibinfo {author} {\bibfnamefont {W.}~\bibnamefont
  {Fox}}, \bibinfo {author} {\bibfnamefont {G.}~\bibnamefont {Fiksel}},
  \bibinfo {author} {\bibfnamefont {A.}~\bibnamefont {Bhattacharjee}}, \bibinfo
  {author} {\bibfnamefont {P.-Y.}\ \bibnamefont {Chang}}, \bibinfo {author}
  {\bibfnamefont {K.}~\bibnamefont {Germaschewski}}, \bibinfo {author}
  {\bibfnamefont {S.~X.}\ \bibnamefont {Hu}}, \ and\ \bibinfo {author}
  {\bibfnamefont {P.~M.}\ \bibnamefont {Nilson}},\ }\bibfield  {title}
  {\enquote {\bibinfo {title} {Filamentation instability of counterstreaming
  laser-driven plasmas},}\ }\href
  {http://dx.doi.org/10.1103/PhysRevLett.111.225002} {\bibfield  {journal}
  {\bibinfo  {journal} {Phys.\ Rev.\ Lett.}\ }\textbf {\bibinfo {volume}
  {111}},\ \bibinfo {pages} {225002} (\bibinfo {year} {2013})}\BibitemShut
  {NoStop}%
\bibitem [{\citenamefont {Niemann}\ \emph {et~al.}(2014)\citenamefont
  {Niemann}, \citenamefont {Gekelman}, \citenamefont {Constantin},
  \citenamefont {Everson}, \citenamefont {Schaeffer}, \citenamefont
  {Bondarenko}, \citenamefont {Clark}, \citenamefont {Winske}, \citenamefont
  {Vincena}, \citenamefont {Van Compernolle},\ and\ \citenamefont
  {Pribyl}}]{Niemann:2014}%
  \BibitemOpen
  \bibfield  {author} {\bibinfo {author} {\bibfnamefont {C.}~\bibnamefont
  {Niemann}}, \bibinfo {author} {\bibfnamefont {W.}~\bibnamefont {Gekelman}},
  \bibinfo {author} {\bibfnamefont {C.~G.}\ \bibnamefont {Constantin}},
  \bibinfo {author} {\bibfnamefont {E.~T.}\ \bibnamefont {Everson}}, \bibinfo
  {author} {\bibfnamefont {D.~B.}\ \bibnamefont {Schaeffer}}, \bibinfo {author}
  {\bibfnamefont {A.~S.}\ \bibnamefont {Bondarenko}}, \bibinfo {author}
  {\bibfnamefont {S.~E.}\ \bibnamefont {Clark}}, \bibinfo {author}
  {\bibfnamefont {D.}~\bibnamefont {Winske}}, \bibinfo {author} {\bibfnamefont
  {S.}~\bibnamefont {Vincena}}, \bibinfo {author} {\bibfnamefont
  {B.}~\bibnamefont {Van Compernolle}}, \ and\ \bibinfo {author}
  {\bibfnamefont {P.}~\bibnamefont {Pribyl}},\ }\bibfield  {title} {\enquote
  {\bibinfo {title} {Observation of collisionless shocks in a large
  current-free laboratory plasma},}\ }\href
  {http://dx.doi.org/10.1002/2014GL061820} {\bibfield  {journal} {\bibinfo
  {journal} {Geophys.\ Res.\ Lett}\ }\textbf {\bibinfo {volume} {41}},\
  \bibinfo {pages} {7413--7418} (\bibinfo {year} {2014})}\BibitemShut {NoStop}%
\bibitem [{\citenamefont {Huntington}\ \emph {et~al.}(2015)\citenamefont
  {Huntington}, \citenamefont {Fiuza}, \citenamefont {Ross}, \citenamefont
  {Zylstra}, \citenamefont {Drake}, \citenamefont {Froula}, \citenamefont
  {Gregori}, \citenamefont {Kugland}, \citenamefont {Kuranz}, \citenamefont
  {Levy}, \citenamefont {Li}, \citenamefont {Meinecke}, \citenamefont {Morita},
  \citenamefont {Petrasso}, \citenamefont {Plechaty}, \citenamefont
  {Remington}, \citenamefont {Ryutov}, \citenamefont {Sakawa}, \citenamefont
  {Spitkovsky}, \citenamefont {Takabe},\ and\ \citenamefont
  {Park}}]{Huntington:2015}%
  \BibitemOpen
  \bibfield  {author} {\bibinfo {author} {\bibfnamefont {C.~M.}\ \bibnamefont
  {Huntington}}, \bibinfo {author} {\bibfnamefont {F.}~\bibnamefont {Fiuza}},
  \bibinfo {author} {\bibfnamefont {J.~S.}\ \bibnamefont {Ross}}, \bibinfo
  {author} {\bibfnamefont {A.~B.}\ \bibnamefont {Zylstra}}, \bibinfo {author}
  {\bibfnamefont {R.~P.}\ \bibnamefont {Drake}}, \bibinfo {author}
  {\bibfnamefont {D.~H.}\ \bibnamefont {Froula}}, \bibinfo {author}
  {\bibfnamefont {G.}~\bibnamefont {Gregori}}, \bibinfo {author} {\bibfnamefont
  {N.~L.}\ \bibnamefont {Kugland}}, \bibinfo {author} {\bibfnamefont {C.~C.}\
  \bibnamefont {Kuranz}}, \bibinfo {author} {\bibfnamefont {M.~C.}\
  \bibnamefont {Levy}}, \bibinfo {author} {\bibfnamefont {C.~K.}\ \bibnamefont
  {Li}}, \bibinfo {author} {\bibfnamefont {J.}~\bibnamefont {Meinecke}},
  \bibinfo {author} {\bibfnamefont {T.}~\bibnamefont {Morita}}, \bibinfo
  {author} {\bibfnamefont {R.}~\bibnamefont {Petrasso}}, \bibinfo {author}
  {\bibfnamefont {C.}~\bibnamefont {Plechaty}}, \bibinfo {author}
  {\bibfnamefont {B.~A.}\ \bibnamefont {Remington}}, \bibinfo {author}
  {\bibfnamefont {D.~D.}\ \bibnamefont {Ryutov}}, \bibinfo {author}
  {\bibfnamefont {Y.}~\bibnamefont {Sakawa}}, \bibinfo {author} {\bibfnamefont
  {A.}~\bibnamefont {Spitkovsky}}, \bibinfo {author} {\bibfnamefont
  {H.}~\bibnamefont {Takabe}}, \ and\ \bibinfo {author} {\bibfnamefont {H.-S.}\
  \bibnamefont {Park}},\ }\bibfield  {title} {\enquote {\bibinfo {title}
  {Observation of magnetic field generation via the weibel instability in
  interpenetrating plasma flows},}\ }\href
  {http://dx.doi.org/10.1038/nphys3178} {\bibfield  {journal} {\bibinfo
  {journal} {Nature\ Phys.}\ }\textbf {\bibinfo {volume} {11}},\ \bibinfo
  {pages} {173--176} (\bibinfo {year} {2015})}\BibitemShut {NoStop}%
\bibitem [{\citenamefont {Schaeffer}\ \emph {et~al.}(2016)\citenamefont
  {Schaeffer}, \citenamefont {Fox}, \citenamefont {Haberberger}, \citenamefont
  {Fiksel}, \citenamefont {Bhattacharjee}, \citenamefont {Barnak},
  \citenamefont {Hu},\ and\ \citenamefont {Germaschewski}}]{Schaeffer:2016}%
  \BibitemOpen
  \bibfield  {author} {\bibinfo {author} {\bibfnamefont {D.~B.}\ \bibnamefont
  {Schaeffer}}, \bibinfo {author} {\bibfnamefont {W.}~\bibnamefont {Fox}},
  \bibinfo {author} {\bibfnamefont {D.}~\bibnamefont {Haberberger}}, \bibinfo
  {author} {\bibfnamefont {G.}~\bibnamefont {Fiksel}}, \bibinfo {author}
  {\bibfnamefont {A.}~\bibnamefont {Bhattacharjee}}, \bibinfo {author}
  {\bibfnamefont {D.}~\bibnamefont {Barnak}}, \bibinfo {author} {\bibfnamefont
  {S.}~\bibnamefont {Hu}}, \ and\ \bibinfo {author} {\bibfnamefont
  {K.}~\bibnamefont {Germaschewski}},\ }\href {http://arxiv.org/abs/1610.06533}
  {\enquote {\bibinfo {title} {Generation and evolution of high-mach number,
  laser-driven magnetized collisionless shocks in the laboratory},}\ }\bibinfo
  {howpublished} {e-print arXiv:physics.plasm-ph/1610.06533} (\bibinfo {year}
  {2016})\BibitemShut {NoStop}%
\bibitem [{\citenamefont {Harnett}\ and\ \citenamefont
  {Winglee}(2000)}]{HarnettWinglee:2000}%
  \BibitemOpen
  \bibfield  {author} {\bibinfo {author} {\bibfnamefont {E.~M.}\ \bibnamefont
  {Harnett}}\ and\ \bibinfo {author} {\bibfnamefont {R.}~\bibnamefont
  {Winglee}},\ }\bibfield  {title} {\enquote {\bibinfo {title} {Two-dimensional
  mhd simulation of the solar wind interaction with magnetic field anomalies on
  the surface of the moon},}\ }\href {http://dx.doi.org/10.1029/2000JA000074}
  {\bibfield  {journal} {\bibinfo  {journal} {J.\ Geophys.\ Res.}\ }\textbf
  {\bibinfo {volume} {105}},\ \bibinfo {pages} {24997---25007} (\bibinfo {year}
  {2000})}\BibitemShut {NoStop}%
\bibitem [{\citenamefont {Harnett}\ and\ \citenamefont
  {Winglee}(2002)}]{HarnettWinglee:2002}%
  \BibitemOpen
  \bibfield  {author} {\bibinfo {author} {\bibfnamefont {E.~M.}\ \bibnamefont
  {Harnett}}\ and\ \bibinfo {author} {\bibfnamefont {R.}~\bibnamefont
  {Winglee}},\ }\bibfield  {title} {\enquote {\bibinfo {title} {2.5d particle
  and mhd simulations of mini-magnetospheres at the moon},}\ }\href
  {http://dx.doi.org/10.1029/2002JA009241} {\bibfield  {journal} {\bibinfo
  {journal} {J.\ Geophys.\ Res.}\ }\textbf {\bibinfo {volume} {107}},\ \bibinfo
  {pages} {SMP 4--1--SMP 4--16} (\bibinfo {year} {2002})}\BibitemShut {NoStop}%
\bibitem [{\citenamefont {Harnett}\ and\ \citenamefont
  {Winglee}(2003)}]{HarnettWinglee:2003}%
  \BibitemOpen
  \bibfield  {author} {\bibinfo {author} {\bibfnamefont {E.~M.}\ \bibnamefont
  {Harnett}}\ and\ \bibinfo {author} {\bibfnamefont {R.}~\bibnamefont
  {Winglee}},\ }\bibfield  {title} {\enquote {\bibinfo {title} {2.5-d fluid
  simulations of the solar wind interacting with multiple dipoles on the
  surface of the moon},}\ }\href {http://dx.doi.org/10.1029/2002JA009617}
  {\bibfield  {journal} {\bibinfo  {journal} {J.\ Geophys.\ Res.}\ }\textbf
  {\bibinfo {volume} {108}} (\bibinfo {year} {2003})}\BibitemShut {NoStop}%
\bibitem [{\citenamefont {Gargat\'e}\ \emph {et~al.}(2014)\citenamefont
  {Gargat\'e}, \citenamefont {Bingham}, \citenamefont {Fonseca}, \citenamefont
  {Bamford}, \citenamefont {Thornton}, \citenamefont {Gibson}, \citenamefont
  {Bradford},\ and\ \citenamefont {Silva}}]{Gargate:2014}%
  \BibitemOpen
  \bibfield  {author} {\bibinfo {author} {\bibfnamefont {L.}~\bibnamefont
  {Gargat\'e}}, \bibinfo {author} {\bibfnamefont {R.}~\bibnamefont {Bingham}},
  \bibinfo {author} {\bibfnamefont {R.~A.}\ \bibnamefont {Fonseca}}, \bibinfo
  {author} {\bibfnamefont {R.~A.}\ \bibnamefont {Bamford}}, \bibinfo {author}
  {\bibfnamefont {A.}~\bibnamefont {Thornton}}, \bibinfo {author}
  {\bibfnamefont {K.}~\bibnamefont {Gibson}}, \bibinfo {author} {\bibfnamefont
  {J.}~\bibnamefont {Bradford}}, \ and\ \bibinfo {author} {\bibfnamefont
  {L.~O.}\ \bibnamefont {Silva}},\ }\bibfield  {title} {\enquote {\bibinfo
  {title} {Sep acceleration in cme driven shocks using a hybrid code},}\ }\href
  {http://dx.doi.org/10.1088/0004-637X/792/1/9} {\bibfield  {journal} {\bibinfo
   {journal} {Astrophys.\ J.}\ }\textbf {\bibinfo {volume} {792}},\ \bibinfo
  {pages} {9} (\bibinfo {year} {2014})}\BibitemShut {NoStop}%
\bibitem [{\citenamefont {Blanco-Cano}, \citenamefont {Omidi},\ and\
  \citenamefont {Russel}(2004)}]{Blanco-Cano:2004}%
  \BibitemOpen
  \bibfield  {author} {\bibinfo {author} {\bibfnamefont {X.}~\bibnamefont
  {Blanco-Cano}}, \bibinfo {author} {\bibfnamefont {N.}~\bibnamefont {Omidi}},
  \ and\ \bibinfo {author} {\bibfnamefont {C.~T.}\ \bibnamefont {Russel}},\
  }\bibfield  {title} {\enquote {\bibinfo {title} {How to make a
  magnetosphere},}\ }\href {http://dx.doi.org/10.1046/j.1468-4004.2003.45314.x}
  {\bibfield  {journal} {\bibinfo  {journal} {Astron.\ Geophys.}\ }\textbf
  {\bibinfo {volume} {45}},\ \bibinfo {pages} {3.14--3.17} (\bibinfo {year}
  {2004})}\BibitemShut {NoStop}%
\bibitem [{\citenamefont {Deca}\ \emph {et~al.}(2014)\citenamefont {Deca},
  \citenamefont {Divin}, \citenamefont {Lapenta}, \citenamefont {Lemb\`ege},
  \citenamefont {Markidis},\ and\ \citenamefont {Hor\'anyi}}]{Deca:2014}%
  \BibitemOpen
  \bibfield  {author} {\bibinfo {author} {\bibfnamefont {J.}~\bibnamefont
  {Deca}}, \bibinfo {author} {\bibfnamefont {A.}~\bibnamefont {Divin}},
  \bibinfo {author} {\bibfnamefont {G.}~\bibnamefont {Lapenta}}, \bibinfo
  {author} {\bibfnamefont {B.}~\bibnamefont {Lemb\`ege}}, \bibinfo {author}
  {\bibfnamefont {S.}~\bibnamefont {Markidis}}, \ and\ \bibinfo {author}
  {\bibfnamefont {M.}~\bibnamefont {Hor\'anyi}},\ }\bibfield  {title} {\enquote
  {\bibinfo {title} {Electromagnetic particle-in-cell simulations of the solar
  wind interaction with lunar magnetic anomalies},}\ }\href
  {http://dx.doi.org/10.1103/PhysRevLett.112.151102} {\bibfield  {journal}
  {\bibinfo  {journal} {Phys.\ Rev.\ Lett.}\ }\textbf {\bibinfo {volume}
  {112}},\ \bibinfo {pages} {151102} (\bibinfo {year} {2014})}\BibitemShut
  {NoStop}%
\bibitem [{\citenamefont {Deca}\ \emph {et~al.}(2015)\citenamefont {Deca},
  \citenamefont {Divin}, \citenamefont {Lemb\`ege}, \citenamefont {Hor\'anyi},
  \citenamefont {Markidis},\ and\ \citenamefont {Lapenta}}]{Deca:2015}%
  \BibitemOpen
  \bibfield  {author} {\bibinfo {author} {\bibfnamefont {J.}~\bibnamefont
  {Deca}}, \bibinfo {author} {\bibfnamefont {A.}~\bibnamefont {Divin}},
  \bibinfo {author} {\bibfnamefont {B.}~\bibnamefont {Lemb\`ege}}, \bibinfo
  {author} {\bibfnamefont {M.}~\bibnamefont {Hor\'anyi}}, \bibinfo {author}
  {\bibfnamefont {S.}~\bibnamefont {Markidis}}, \ and\ \bibinfo {author}
  {\bibfnamefont {G.}~\bibnamefont {Lapenta}},\ }\bibfield  {title} {\enquote
  {\bibinfo {title} {General mechanism and dynamics of the solar wind
  interaction with lunar magnetic anomalies from 3-d pic simulations},}\ }\href
  {http://dx.doi.org/10.1002/2015JA021070} {\bibfield  {journal} {\bibinfo
  {journal} {J.\ Geophys.\ Res. Space\ Physics}\ }\textbf {\bibinfo {volume}
  {120}},\ \bibinfo {pages} {6443--6463} (\bibinfo {year} {2015})}\BibitemShut
  {NoStop}%
\bibitem [{\citenamefont {Ashida}\ \emph {et~al.}(2014)\citenamefont {Ashida},
  \citenamefont {Usui}, \citenamefont {Shinohara}, \citenamefont {Nakamura},
  \citenamefont {Funaki}, \citenamefont {Miyake},\ and\ \citenamefont
  {Yamakawa}}]{Ashida:2014}%
  \BibitemOpen
  \bibfield  {author} {\bibinfo {author} {\bibfnamefont {Y.}~\bibnamefont
  {Ashida}}, \bibinfo {author} {\bibfnamefont {H.}~\bibnamefont {Usui}},
  \bibinfo {author} {\bibfnamefont {I.}~\bibnamefont {Shinohara}}, \bibinfo
  {author} {\bibfnamefont {M.}~\bibnamefont {Nakamura}}, \bibinfo {author}
  {\bibfnamefont {I.}~\bibnamefont {Funaki}}, \bibinfo {author} {\bibfnamefont
  {Y.}~\bibnamefont {Miyake}}, \ and\ \bibinfo {author} {\bibfnamefont
  {H.}~\bibnamefont {Yamakawa}},\ }\bibfield  {title} {\enquote {\bibinfo
  {title} {Full kinetic simulations of plasma flow interactions with meso- and
  microscale magnetic dipoles},}\ }\href {http://dx.doi.org/10.1063/1.4904303}
  {\bibfield  {journal} {\bibinfo  {journal} {Phys.\ Plasmas}\ }\textbf
  {\bibinfo {volume} {21}},\ \bibinfo {pages} {122903} (\bibinfo {year}
  {2014})}\BibitemShut {NoStop}%
\bibitem [{\citenamefont {Bamford}\ \emph {et~al.}(2016)\citenamefont
  {Bamford}, \citenamefont {Alves}, \citenamefont {Cruz}, \citenamefont
  {Kellett}, \citenamefont {Fonseca}, \citenamefont {Silva}, \citenamefont
  {Trines}, \citenamefont {Halekas}, \citenamefont {Kramer}, \citenamefont
  {Harnett}, \citenamefont {Cairns},\ and\ \citenamefont
  {Bingham}}]{Bamford:2016}%
  \BibitemOpen
  \bibfield  {author} {\bibinfo {author} {\bibfnamefont {R.~A.}\ \bibnamefont
  {Bamford}}, \bibinfo {author} {\bibfnamefont {E.~P.}\ \bibnamefont {Alves}},
  \bibinfo {author} {\bibfnamefont {F.}~\bibnamefont {Cruz}}, \bibinfo {author}
  {\bibfnamefont {B.~J.}\ \bibnamefont {Kellett}}, \bibinfo {author}
  {\bibfnamefont {R.~A.}\ \bibnamefont {Fonseca}}, \bibinfo {author}
  {\bibfnamefont {L.~O.}\ \bibnamefont {Silva}}, \bibinfo {author}
  {\bibfnamefont {R.~M. G.~M.}\ \bibnamefont {Trines}}, \bibinfo {author}
  {\bibfnamefont {J.~S.}\ \bibnamefont {Halekas}}, \bibinfo {author}
  {\bibfnamefont {G.}~\bibnamefont {Kramer}}, \bibinfo {author} {\bibfnamefont
  {E.}~\bibnamefont {Harnett}}, \bibinfo {author} {\bibfnamefont {R.~A.}\
  \bibnamefont {Cairns}}, \ and\ \bibinfo {author} {\bibfnamefont
  {R.}~\bibnamefont {Bingham}},\ }\bibfield  {title} {\enquote {\bibinfo
  {title} {3d {PIC} simulations of collisionless shocks at lunar magnetic
  anomalies and their role in forming lunar swirls},}\ }\href
  {http://dx.doi.org/10.3847/0004-637X/830/2/146} {\bibfield  {journal}
  {\bibinfo  {journal} {Astrophys.\ J.}\ }\textbf {\bibinfo {volume} {830}},\
  \bibinfo {pages} {146} (\bibinfo {year} {2016})}\BibitemShut {NoStop}%
\bibitem [{\citenamefont {Fonseca}\ \emph {et~al.}(2012)\citenamefont
  {Fonseca}, \citenamefont {Silva}, \citenamefont {Tsung}, \citenamefont
  {Decyk}, \citenamefont {Lu}, \citenamefont {Ren}, \citenamefont {Mori},
  \citenamefont {Deng}, \citenamefont {Lee}, \citenamefont {Katsouleas},\ and\
  \citenamefont {Adam}}]{Fonseca:2012}%
  \BibitemOpen
  \bibfield  {author} {\bibinfo {author} {\bibfnamefont {R.~A.}\ \bibnamefont
  {Fonseca}}, \bibinfo {author} {\bibfnamefont {L.~O.}\ \bibnamefont {Silva}},
  \bibinfo {author} {\bibfnamefont {F.~S.}\ \bibnamefont {Tsung}}, \bibinfo
  {author} {\bibfnamefont {V.~K.}\ \bibnamefont {Decyk}}, \bibinfo {author}
  {\bibfnamefont {W.}~\bibnamefont {Lu}}, \bibinfo {author} {\bibfnamefont
  {C.}~\bibnamefont {Ren}}, \bibinfo {author} {\bibfnamefont {W.~B.}\
  \bibnamefont {Mori}}, \bibinfo {author} {\bibfnamefont {S.}~\bibnamefont
  {Deng}}, \bibinfo {author} {\bibfnamefont {S.}~\bibnamefont {Lee}}, \bibinfo
  {author} {\bibfnamefont {T.}~\bibnamefont {Katsouleas}}, \ and\ \bibinfo
  {author} {\bibfnamefont {J.~C.}\ \bibnamefont {Adam}},\ }\bibfield  {title}
  {\enquote {\bibinfo {title} {{OSIRIS}: A three-dimensional, fully
  relativistic particle in cell code for modeling plasma based accelerators},}\
  }in\ \href {http://dx.doi.org/10.1007/3-540-47789-6\_36} {\emph {\bibinfo
  {booktitle} {Computational Science — ICCS 2002}}},\ \bibinfo {series}
  {Lecture Notes in Computer Science}, Vol.\ \bibinfo {volume} {2331}\
  (\bibinfo  {publisher} {Springer Berlin Heidelberg},\ \bibinfo {year}
  {2012})\ pp.\ \bibinfo {pages} {342--351}\BibitemShut {NoStop}%
\bibitem [{\citenamefont {Fonseca}\ \emph {et~al.}(2013)\citenamefont
  {Fonseca}, \citenamefont {Vieira}, \citenamefont {Fiuza}, \citenamefont
  {Davidson}, \citenamefont {Tsung}, \citenamefont {Mori},\ and\ \citenamefont
  {Silva}}]{Fonseca:2013}%
  \BibitemOpen
  \bibfield  {author} {\bibinfo {author} {\bibfnamefont {R.~A.}\ \bibnamefont
  {Fonseca}}, \bibinfo {author} {\bibfnamefont {J.}~\bibnamefont {Vieira}},
  \bibinfo {author} {\bibfnamefont {F.}~\bibnamefont {Fiuza}}, \bibinfo
  {author} {\bibfnamefont {A.}~\bibnamefont {Davidson}}, \bibinfo {author}
  {\bibfnamefont {F.~S.}\ \bibnamefont {Tsung}}, \bibinfo {author}
  {\bibfnamefont {W.~B.}\ \bibnamefont {Mori}}, \ and\ \bibinfo {author}
  {\bibfnamefont {L.~O.}\ \bibnamefont {Silva}},\ }\bibfield  {title} {\enquote
  {\bibinfo {title} {Exploiting multi-scale parallelism for large scale
  numerical modelling of laser wakefield accelerators},}\ }\href
  {http://dx.doi.org/10.1088/0741-3335/55/12/124011} {\bibfield  {journal}
  {\bibinfo  {journal} {Plasma\ Phys.\ Controlled\ Fusion}\ }\textbf {\bibinfo
  {volume} {55}},\ \bibinfo {pages} {124011} (\bibinfo {year}
  {2013})}\BibitemShut {NoStop}%
\bibitem [{\citenamefont {McBride}\ \emph {et~al.}(1972)\citenamefont
  {McBride}, \citenamefont {Ott}, \citenamefont {Boris},\ and\ \citenamefont
  {Orens}}]{McBride:1972}%
  \BibitemOpen
  \bibfield  {author} {\bibinfo {author} {\bibfnamefont {J.~B.}\ \bibnamefont
  {McBride}}, \bibinfo {author} {\bibfnamefont {E.}~\bibnamefont {Ott}},
  \bibinfo {author} {\bibfnamefont {J.~P.}\ \bibnamefont {Boris}}, \ and\
  \bibinfo {author} {\bibfnamefont {J.~H.}\ \bibnamefont {Orens}},\ }\bibfield
  {title} {\enquote {\bibinfo {title} {Theory and simulation of turbulent
  heating by the modified two-stream instability},}\ }\href
  {http://dx.doi.org/10.1063/1.1693881} {\bibfield  {journal} {\bibinfo
  {journal} {Phys.\ Fluids}\ }\textbf {\bibinfo {volume} {15}},\ \bibinfo
  {pages} {2367--2383} (\bibinfo {year} {1972})}\BibitemShut {NoStop}%
\bibitem [{\citenamefont {Parker}(1979)}]{Parker:1979}%
  \BibitemOpen
  \bibfield  {author} {\bibinfo {author} {\bibfnamefont {E.~N.}\ \bibnamefont
  {Parker}},\ }\href@noop {} {\emph {\bibinfo {title} {Cosmical magnetic
  fields: Their origin and their activity}}}\ (\bibinfo  {publisher} {Oxford
  University Press, Clarendon Press},\ \bibinfo {address} {Oxford-New York},\
  \bibinfo {year} {1979})\BibitemShut {NoStop}%
\bibitem [{\citenamefont {Priest}\ and\ \citenamefont
  {Forbes}(2010)}]{PriestForbes:2010}%
  \BibitemOpen
  \bibfield  {author} {\bibinfo {author} {\bibfnamefont {E.~R.}\ \bibnamefont
  {Priest}}\ and\ \bibinfo {author} {\bibfnamefont {T.}~\bibnamefont
  {Forbes}},\ }\href@noop {} {\emph {\bibinfo {title} {Magnetic Reconnection:
  MHD theory and applications}}},\ \bibinfo {edition} {1st}\ ed.\ (\bibinfo
  {publisher} {Cambridge University Press},\ \bibinfo {address} {Cambridge-New
  York-Melbourne-Madrid},\ \bibinfo {year} {2010})\BibitemShut {NoStop}%
\bibitem [{\citenamefont {Yamada}, \citenamefont {Kulsrud},\ and\ \citenamefont
  {Ji}(2010)}]{Yamada:2010}%
  \BibitemOpen
  \bibfield  {author} {\bibinfo {author} {\bibfnamefont {M.}~\bibnamefont
  {Yamada}}, \bibinfo {author} {\bibfnamefont {R.}~\bibnamefont {Kulsrud}}, \
  and\ \bibinfo {author} {\bibfnamefont {H.}~\bibnamefont {Ji}},\ }\bibfield
  {title} {\enquote {\bibinfo {title} {Magnetic reconnection},}\ }\href
  {http://dx.doi.org/10.1103/RevModPhys.82.603} {\bibfield  {journal} {\bibinfo
   {journal} {Rev.\ Mod.\ Phys.}\ }\textbf {\bibinfo {volume} {82}},\ \bibinfo
  {pages} {603--664} (\bibinfo {year} {2010})}\BibitemShut {NoStop}%
\bibitem [{\citenamefont {Markidis}\ \emph {et~al.}(2012)\citenamefont
  {Markidis}, \citenamefont {Henri}, \citenamefont {Lapenta}, \citenamefont
  {Divin}, \citenamefont {Goldman}, \citenamefont {Newman},\ and\ \citenamefont
  {Eriksson}}]{Markidis:2012}%
  \BibitemOpen
  \bibfield  {author} {\bibinfo {author} {\bibfnamefont {S.}~\bibnamefont
  {Markidis}}, \bibinfo {author} {\bibfnamefont {P.}~\bibnamefont {Henri}},
  \bibinfo {author} {\bibfnamefont {G.}~\bibnamefont {Lapenta}}, \bibinfo
  {author} {\bibfnamefont {A.}~\bibnamefont {Divin}}, \bibinfo {author}
  {\bibfnamefont {M.~V.}\ \bibnamefont {Goldman}}, \bibinfo {author}
  {\bibfnamefont {D.}~\bibnamefont {Newman}}, \ and\ \bibinfo {author}
  {\bibfnamefont {S.}~\bibnamefont {Eriksson}},\ }\bibfield  {title} {\enquote
  {\bibinfo {title} {Collisionless magnetic reconnection in a plasmoid
  chain},}\ }\href {http://dx.doi.org/10.5194/npg-19-145-2012} {\bibfield
  {journal} {\bibinfo  {journal} {Nonlin.\ Processes\ Geophys.}\ }\textbf
  {\bibinfo {volume} {19}},\ \bibinfo {pages} {145--153} (\bibinfo {year}
  {2012})}\BibitemShut {NoStop}%
\bibitem [{\citenamefont {Loureiro}, \citenamefont {Schekochihin},\ and\
  \citenamefont {Cowley}(2007)}]{Loureiro:2007}%
  \BibitemOpen
  \bibfield  {author} {\bibinfo {author} {\bibfnamefont {N.~F.}\ \bibnamefont
  {Loureiro}}, \bibinfo {author} {\bibfnamefont {A.~A.}\ \bibnamefont
  {Schekochihin}}, \ and\ \bibinfo {author} {\bibfnamefont {S.~C.}\
  \bibnamefont {Cowley}},\ }\bibfield  {title} {\enquote {\bibinfo {title}
  {Instability of current sheets and formation of plasmoid chains},}\ }\href
  {http://dx.doi.org/10.1063/1.2783986} {\bibfield  {journal} {\bibinfo
  {journal} {Phys.\ Plasmas}\ }\textbf {\bibinfo {volume} {14}},\ \bibinfo
  {pages} {100703} (\bibinfo {year} {2007})}\BibitemShut {NoStop}%
\bibitem [{\citenamefont {Samtaney}\ \emph {et~al.}(2009)\citenamefont
  {Samtaney}, \citenamefont {Loureiro}, \citenamefont {Uzdensky}, \citenamefont
  {Schekochihin},\ and\ \citenamefont {Cowley}}]{Samtaney:2009}%
  \BibitemOpen
  \bibfield  {author} {\bibinfo {author} {\bibfnamefont {R.}~\bibnamefont
  {Samtaney}}, \bibinfo {author} {\bibfnamefont {N.~F.}\ \bibnamefont
  {Loureiro}}, \bibinfo {author} {\bibfnamefont {D.~A.}\ \bibnamefont
  {Uzdensky}}, \bibinfo {author} {\bibfnamefont {A.~A.}\ \bibnamefont
  {Schekochihin}}, \ and\ \bibinfo {author} {\bibfnamefont {S.~C.}\
  \bibnamefont {Cowley}},\ }\bibfield  {title} {\enquote {\bibinfo {title}
  {Formation of plasmoid chains in magnetic reconnection},}\ }\href
  {http://dx.doi.org/10.1103/PhysRevLett.103.105004} {\bibfield  {journal}
  {\bibinfo  {journal} {Phys.\ Rev.\ Lett.}\ }\textbf {\bibinfo {volume}
  {103}},\ \bibinfo {pages} {105004} (\bibinfo {year} {2009})}\BibitemShut
  {NoStop}%
\bibitem [{\citenamefont {Gekelman}\ \emph {et~al.}(1991)\citenamefont
  {Gekelman}, \citenamefont {Pfister}, \citenamefont {Lucky}, \citenamefont
  {Bamber}, \citenamefont {Leneman},\ and\ \citenamefont
  {Maggs}}]{Gekelman:1991}%
  \BibitemOpen
  \bibfield  {author} {\bibinfo {author} {\bibfnamefont {W.}~\bibnamefont
  {Gekelman}}, \bibinfo {author} {\bibfnamefont {H.}~\bibnamefont {Pfister}},
  \bibinfo {author} {\bibfnamefont {Z.}~\bibnamefont {Lucky}}, \bibinfo
  {author} {\bibfnamefont {J.}~\bibnamefont {Bamber}}, \bibinfo {author}
  {\bibfnamefont {D.}~\bibnamefont {Leneman}}, \ and\ \bibinfo {author}
  {\bibfnamefont {J.}~\bibnamefont {Maggs}},\ }\bibfield  {title} {\enquote
  {\bibinfo {title} {Design, construction, and properties of the large plasma
  research device−the lapd at ucla},}\ }\href
  {http://dx.doi.org/10.1063/1.1142175} {\bibfield  {journal} {\bibinfo
  {journal} {Rev.\ Sci.\ Instrum.}\ }\textbf {\bibinfo {volume} {62}},\
  \bibinfo {pages} {2875--2883} (\bibinfo {year} {1991})}\BibitemShut {NoStop}%
\bibitem [{\citenamefont {Niemann}\ \emph {et~al.}(2012)\citenamefont
  {Niemann}, \citenamefont {Constantin}, \citenamefont {Schaeffer},
  \citenamefont {Tauschwitz}, \citenamefont {Weiland}, \citenamefont {Lucky},
  \citenamefont {Gekelman}, \citenamefont {Everson},\ and\ \citenamefont
  {Winske}}]{Niemann:2012}%
  \BibitemOpen
  \bibfield  {author} {\bibinfo {author} {\bibfnamefont {C.}~\bibnamefont
  {Niemann}}, \bibinfo {author} {\bibfnamefont {C.~G.}\ \bibnamefont
  {Constantin}}, \bibinfo {author} {\bibfnamefont {D.~B.}\ \bibnamefont
  {Schaeffer}}, \bibinfo {author} {\bibfnamefont {A.}~\bibnamefont
  {Tauschwitz}}, \bibinfo {author} {\bibfnamefont {T.}~\bibnamefont {Weiland}},
  \bibinfo {author} {\bibfnamefont {Z.}~\bibnamefont {Lucky}}, \bibinfo
  {author} {\bibfnamefont {W.}~\bibnamefont {Gekelman}}, \bibinfo {author}
  {\bibfnamefont {E.~T.}\ \bibnamefont {Everson}}, \ and\ \bibinfo {author}
  {\bibfnamefont {D.}~\bibnamefont {Winske}},\ }\bibfield  {title} {\enquote
  {\bibinfo {title} {High-energy nd:glass laser facility for collisionless
  laboratory astrophysics},}\ }\href
  {http://dx.doi.org/10.1088/1748-0221/7/03/P03010} {\bibfield  {journal}
  {\bibinfo  {journal} {J.\ Instrum.}\ }\textbf {\bibinfo {volume} {7}},\
  \bibinfo {pages} {P03010} (\bibinfo {year} {2012})}\BibitemShut {NoStop}%
\bibitem [{\citenamefont {Waxer}\ \emph {et~al.}(2005)\citenamefont {Waxer},
  \citenamefont {Maywar}, \citenamefont {Kelly}, \citenamefont {Kessler},
  \citenamefont {Kruschwitz}, \citenamefont {Loucks}, \citenamefont {McCrory},
  \citenamefont {Meyerhofer}, \citenamefont {Morse}, \citenamefont {Stoeckl},\
  and\ \citenamefont {Zuegel}}]{Waxer:2005}%
  \BibitemOpen
  \bibfield  {author} {\bibinfo {author} {\bibfnamefont {L.~J.}\ \bibnamefont
  {Waxer}}, \bibinfo {author} {\bibfnamefont {D.~N.}\ \bibnamefont {Maywar}},
  \bibinfo {author} {\bibfnamefont {J.~H.}\ \bibnamefont {Kelly}}, \bibinfo
  {author} {\bibfnamefont {T.~J.}\ \bibnamefont {Kessler}}, \bibinfo {author}
  {\bibfnamefont {B.~E.}\ \bibnamefont {Kruschwitz}}, \bibinfo {author}
  {\bibfnamefont {S.~J.}\ \bibnamefont {Loucks}}, \bibinfo {author}
  {\bibfnamefont {R.~L.}\ \bibnamefont {McCrory}}, \bibinfo {author}
  {\bibfnamefont {D.~D.}\ \bibnamefont {Meyerhofer}}, \bibinfo {author}
  {\bibfnamefont {S.~F.~B.}\ \bibnamefont {Morse}}, \bibinfo {author}
  {\bibfnamefont {C.}~\bibnamefont {Stoeckl}}, \ and\ \bibinfo {author}
  {\bibfnamefont {J.~D.}\ \bibnamefont {Zuegel}},\ }\bibfield  {title}
  {\enquote {\bibinfo {title} {High-energy petawatt capability for the omega
  laser},}\ }\href {http://dx.doi.org/10.1364/OPN.16.7.000030} {\bibfield
  {journal} {\bibinfo  {journal} {Opt.\ Photon.\ News}\ }\textbf {\bibinfo
  {volume} {16}},\ \bibinfo {pages} {30--36} (\bibinfo {year}
  {2005})}\BibitemShut {NoStop}%
\bibitem [{\citenamefont {Constantin}\ \emph {et~al.}(2009)\citenamefont
  {Constantin}, \citenamefont {Gekelman}, \citenamefont {Pribyl}, \citenamefont
  {Everson}, \citenamefont {Schaeffer}, \citenamefont {Kugland}, \citenamefont
  {Presura}, \citenamefont {Neff}, \citenamefont {Plechaty}, \citenamefont
  {Vincena}, \citenamefont {Collette}, \citenamefont {Tripathi}, \citenamefont
  {Muniz},\ and\ \citenamefont {Niemann}}]{Constantin:2009}%
  \BibitemOpen
  \bibfield  {author} {\bibinfo {author} {\bibfnamefont {C.}~\bibnamefont
  {Constantin}}, \bibinfo {author} {\bibfnamefont {W.}~\bibnamefont
  {Gekelman}}, \bibinfo {author} {\bibfnamefont {P.}~\bibnamefont {Pribyl}},
  \bibinfo {author} {\bibfnamefont {E.}~\bibnamefont {Everson}}, \bibinfo
  {author} {\bibfnamefont {D.~B.}\ \bibnamefont {Schaeffer}}, \bibinfo {author}
  {\bibfnamefont {N.}~\bibnamefont {Kugland}}, \bibinfo {author} {\bibfnamefont
  {R.}~\bibnamefont {Presura}}, \bibinfo {author} {\bibfnamefont
  {S.}~\bibnamefont {Neff}}, \bibinfo {author} {\bibfnamefont {C.}~\bibnamefont
  {Plechaty}}, \bibinfo {author} {\bibfnamefont {S.}~\bibnamefont {Vincena}},
  \bibinfo {author} {\bibfnamefont {A.}~\bibnamefont {Collette}}, \bibinfo
  {author} {\bibfnamefont {S.}~\bibnamefont {Tripathi}}, \bibinfo {author}
  {\bibfnamefont {M.~V.}\ \bibnamefont {Muniz}}, \ and\ \bibinfo {author}
  {\bibfnamefont {C.}~\bibnamefont {Niemann}},\ }\bibfield  {title} {\enquote
  {\bibinfo {title} {Collisionless interaction of an energetic laser produced
  plasma with a large magnetoplasma},}\ }\href
  {http://dx.doi.org/10.1007/s10509-009-0012-z} {\bibfield  {journal} {\bibinfo
   {journal} {Astrophys.\ Space\ Sci.}\ }\textbf {\bibinfo {volume} {322}},\
  \bibinfo {pages} {155--159} (\bibinfo {year} {2009})}\BibitemShut {NoStop}%
\bibitem [{\citenamefont {Niemann}\ \emph {et~al.}(2013)\citenamefont
  {Niemann}, \citenamefont {Gekelman}, \citenamefont {Constantin},
  \citenamefont {Everson}, \citenamefont {Schaeffer}, \citenamefont {Clark},
  \citenamefont {Winske}, \citenamefont {Zylstra}, \citenamefont {Pribyl},
  \citenamefont {Tripathi}, \citenamefont {Larson}, \citenamefont {Glenzer},\
  and\ \citenamefont {Bondarenko}}]{Niemann:2013}%
  \BibitemOpen
  \bibfield  {author} {\bibinfo {author} {\bibfnamefont {C.}~\bibnamefont
  {Niemann}}, \bibinfo {author} {\bibfnamefont {W.}~\bibnamefont {Gekelman}},
  \bibinfo {author} {\bibfnamefont {C.~G.}\ \bibnamefont {Constantin}},
  \bibinfo {author} {\bibfnamefont {E.~T.}\ \bibnamefont {Everson}}, \bibinfo
  {author} {\bibfnamefont {D.~B.}\ \bibnamefont {Schaeffer}}, \bibinfo {author}
  {\bibfnamefont {S.~E.}\ \bibnamefont {Clark}}, \bibinfo {author}
  {\bibfnamefont {D.}~\bibnamefont {Winske}}, \bibinfo {author} {\bibfnamefont
  {A.~B.}\ \bibnamefont {Zylstra}}, \bibinfo {author} {\bibfnamefont
  {P.}~\bibnamefont {Pribyl}}, \bibinfo {author} {\bibfnamefont {S.~K.~P.}\
  \bibnamefont {Tripathi}}, \bibinfo {author} {\bibfnamefont {D.}~\bibnamefont
  {Larson}}, \bibinfo {author} {\bibfnamefont {S.~H.}\ \bibnamefont {Glenzer}},
  \ and\ \bibinfo {author} {\bibfnamefont {A.~S.}\ \bibnamefont {Bondarenko}},\
  }\bibfield  {title} {\enquote {\bibinfo {title} {Dynamics of exploding
  plasmas in a large magnetized plasma},}\ }\href
  {http://dx.doi.org/10.1063/1.4773911} {\bibfield  {journal} {\bibinfo
  {journal} {Phys.\ Plasmas}\ }\textbf {\bibinfo {volume} {20}},\ \bibinfo
  {pages} {012108} (\bibinfo {year} {2013})}\BibitemShut {NoStop}%
\bibitem [{\citenamefont {Schaeffer}\ \emph {et~al.}(2012)\citenamefont
  {Schaeffer}, \citenamefont {Everson}, \citenamefont {Winske}, \citenamefont
  {Constantin}, \citenamefont {Bondarenko}, \citenamefont {Morton},
  \citenamefont {Flippo}, \citenamefont {Montgomery}, \citenamefont
  {Gaillard},\ and\ \citenamefont {Niemann}}]{Schaeffer:2012}%
  \BibitemOpen
  \bibfield  {author} {\bibinfo {author} {\bibfnamefont {D.~B.}\ \bibnamefont
  {Schaeffer}}, \bibinfo {author} {\bibfnamefont {E.~T.}\ \bibnamefont
  {Everson}}, \bibinfo {author} {\bibfnamefont {D.}~\bibnamefont {Winske}},
  \bibinfo {author} {\bibfnamefont {C.~G.}\ \bibnamefont {Constantin}},
  \bibinfo {author} {\bibfnamefont {A.~S.}\ \bibnamefont {Bondarenko}},
  \bibinfo {author} {\bibfnamefont {L.~A.}\ \bibnamefont {Morton}}, \bibinfo
  {author} {\bibfnamefont {K.~A.}\ \bibnamefont {Flippo}}, \bibinfo {author}
  {\bibfnamefont {D.~S.}\ \bibnamefont {Montgomery}}, \bibinfo {author}
  {\bibfnamefont {S.~A.}\ \bibnamefont {Gaillard}}, \ and\ \bibinfo {author}
  {\bibfnamefont {C.}~\bibnamefont {Niemann}},\ }\bibfield  {title} {\enquote
  {\bibinfo {title} {Generation of magnetized collisionless shocks by a novel,
  laser-driven magnetic piston},}\ }\href {http://dx.doi.org/10.1063/1.4736846}
  {\bibfield  {journal} {\bibinfo  {journal} {Phys.\ Plasmas}\ }\textbf
  {\bibinfo {volume} {19}},\ \bibinfo {pages} {070702} (\bibinfo {year}
  {2012})}\BibitemShut {NoStop}%
\end{thebibliography}%

\end{document}